\documentclass{kapproc} 

\kluwerbib

\def\lesssim{\mathrel{\hbox{\rlap{\hbox{\lower4pt\hbox{$\sim$}}}\hbox{$<$}}}}

\def\PRL{{\em Phys. Rev. Lett.}\ }

\def\PRD{{\em Phys. Rev.} D}
\def\PRC{{\em Phys. Rev.} C}

\def\apj{{\em Ap.~J.}\ }

\def\aap{{\em Astr.~Ap.}\ }

\newcommand{\anu}{\bar{\nu}}
\newcommand{\aenu}{\bar{\nu}_e}
\newcommand{\enu}{\nu_e}
\newcommand{\unu}{\nu_\mu}
\newcommand{\aunu}{\bar{\nu}_\mu}

\newcommand{\vep}{\epsilon}
\newcommand{\epnu}{\varepsilon_\nu}

\newcommand{\avemu}{\langle\mu_{\nu_i}\rangle}
\newcommand{\aveamu}{\langle\mu_{\bar{\nu}_i}\rangle}
\newcommand{\avemut}{\langle\mu_{\nu_i}^2\rangle}

\newcommand{\f}{{\cal{F}}}
\newcommand{\sinw}{\sin^2\theta_W}

\newcommand{\beq}{\begin{equation}} 
\newcommand{\eeq}{\end{equation}} 
\newcommand{\beqa}{\begin{eqnarray}} 
\newcommand{\eeqa}{\end{eqnarray}} 
    
\newcommand{\pr}{^\prime}    
\newcommand{\righta}{\rightarrow}
\newcommand{\sig}{\sigma}

\begin{document}

\articletitle{Neutrino-Matter Interaction Rates in Supernovae}

\articlesubtitle{The Essential Microphysics of Core Collapse}

\author{Adam Burrows}
\affil{Steward Observatory, 
The University of Arizona, Tucson, AZ 85721 \\}
\email{burrows@zenith.as.arizona.edu}

\author{Todd A. Thompson\footnote{Hubble Fellow}}
\affil{Astronomy Department and Theoretical Astrophysics Center, \\
601 Campbell Hall, The University of California, Berkeley, CA 94720}
\email{thomp@astro.berkeley.edu}

\begin{abstract}
Neutrino-matter interaction rates are central to
the core collapse phenomenon and, perhaps, to the viability
of the mechanism of core-collapse supernova explosions.  In
this paper we catalog and discuss the major neutrino scattering,
absorption, and production processes that together influence
the outcome of core collapse and the cooling of protoneutron
stars.  These are the essential inputs into the codes used
to simulate the supernova phenomenon and an understanding of
these processes is a prerequisite to continuing progress
in supernova theory.
\end{abstract}


\section{Introduction}

One of the key insights of the 20th Century was that most of the elements of nature are
created by nuclear processes in stars.   Supernova explosions are one major means by which these elements
are injected into the interstellar medium and, hence, into subsequent generations of stars.  Therefore, supernovae are
central to the chemical evolution and progressive enrichment of the universe. 
Core-collapse supernova explosions signal the death of a massive star
and are some of the most majestic and awe-inspiring events in the cosmos. However, to fully understand
the role of supernovae in the grand synthesis of creation, one must have a firm handle on the nuclear
data.  In particular, understanding the origin of iron-peak, r-process, and rp-process elements hinges upon  
improvements in our knowledge of the properties of exotic nuclei, some of which are far from the 
valley of beta stability.  

The mechanism of core-collapse supernovae is thought to depend upon the transfer of energy from the core to the
mantle of the inner regions of a massive star after it becomes unstable to collapse.  Neutrinos seem to
be the mediators of this energy transfer.  In order to understand this coupling and the role of neutrinos
in supernova explosions, one needs to master the particulars of the neutrino-matter scattering, production,
and absorption rates.  Since recently there has been some progress in understanding the associated microphysics,
it is fitting to summarize the neutrino-matter cross sections and the production rates of neutrinos in the core-collapse
context.  To this end, we have assembled here a short pr\'ecis of many of the relevant processes and physics.
This contribution does not attempt to explain the hydrodynamics of supernova explosions, but does try
to present the relevant neutrino processes that play a role.  For the former, the reader is referred to
\cite{bhf_1995} and \cite{nature}.  

In \S\ref{stimabs}, we present 
a physical derivation of stimulated absorption (the Fermionic correlate to stimulated emission) and then in \S\ref{cross6}
we present the basic cross sections.  In \S\ref{section:inelastic}, we discuss
the inelastic neutrino-electron and neutrino-nucleon scattering processes and energy redistribution.  This is followed with 
a discussion of the alternate, more powerful formalism, for determining differential
interaction rates in the many-body context and for handling redistribution, namely that of dynamical structure factors.    
Both the non-interacting and the interacting (in the context of a simple nuclear model)
cases are explored, as well as collective excitations of the medium.
Source terms for electron-positron annihilation (\S\ref{eplus}), 
neutrino-anti-neutrino annihilation (\S\ref{paira}; Janka 1991),
and nucleon-nucleon bremsstrahlung (\S\ref{bremsst}) cap off our review of the major
processes of relevance in core-collapse simulations.
Nucleon--nucleon bremsstrahlung can compete with pair annihilation as a source for $\nu_{\mu}$, $\bar{\nu}_{\mu}$,
$\nu_{\tau}$, and $\bar{\nu}_{\tau}$ neutrinos.
Clearly, a mastery of these neutrino-matter processes is a prerequisite
for progress in supernova theory and it is in that spirit that we have assembled 
this review.

\section{Stimulated Absorption}
\label{stimabs}

The concept of stimulated emission for photons is well understood and studied, but
the corresponding concept of stimulated {\it absorption} for neutrinos is not so
well appreciated.  This may be because its simple origin in Fermi blocking and the
Pauli exclusion principle in the context of {\it net} emission is not often explained.
The {\it net} emission of a neutrino is simply the difference between the emissivity and
the absorption of the medium:
\beq
{\cal {J}}_{net}=\eta_\nu-\kappa_aI_\nu\,\, .
\label{netemission}
\eeq
All absorption processes involving fermions will be inhibited by Pauli blocking due to
final--state occupancy.  Hence, $\eta_\nu$ in eq. (\ref{netemission}) includes a blocking term,
$(1-\f_{\nu})$(Bruenn 1985). $\f_{\nu}$ is the invariant distribution function
for the neutrino, whether or not it is in chemical equilibrium.

We can derive stimulated absorption using Fermi's Golden rule.  For example, the net collision term
for the process, $\nu_en\leftrightarrow e^-p$, is:
\beqa
{\cal C}_{\nu_e n\leftrightarrow e^- p}&=&
\int\frac{d^3\vec{p}_{\nu_e}}{(2\pi)^3 }
\int\frac{d^3\vec{p}_n}{(2\pi)^3 }
\int\frac{d^3\vec{p}_p}{(2\pi)^3 }
\int\frac{d^3\vec{p}_e}{(2\pi)^3 }\,
\left(\sum_{s}{|{\cal{M}}|}^2\right) \nonumber \\
&& \nonumber \\
&\times&
\Xi(\nu_en\leftrightarrow e^-p)\,\,(2\pi)^4\,\delta^4({\bf p}_{\nu_e}+{\bf
p}_n-{\bf p}_p-{\bf p}_e)
\,\, ,
\label{rate01}
\eeqa
where ${\bf p}$ is a four-vector and
\beq
\Xi(\nu_en\leftrightarrow
e^-p)=\f_{\nu_e}\f_n(1-\f_e)(1-\f_p)-\f_e\f_p(1-\f_n)(1-\f_{\nu_e})\,\, .
\label{blocks}
\eeq
The final--state blocking terms in eq. (\ref{blocks}) are manifest, in particular that for the $\nu_e$ neutrino.
Algebraic manipulations convert $\Xi(\nu_en\leftrightarrow e^-p)$ in eq. (\ref{blocks}) into:
\beqa
\Xi(\nu_en\leftrightarrow e^-p)&=&
\f_n(1-\f_e)(1-\f_p)\left[\frac{\f_{\nu_e}^{\rm eq}}{1-\f_{\nu_e}^{\rm eq}}(1-\f_{\nu_e})-\f_{\nu_e}\right]
\nonumber \\
&=&\frac{\f_n(1-\f_e)(1-\f_p)}{1-\f_{\nu_e}\pr}\left[\f_{\nu_e}^{\rm eq}-\f_{\nu_e}\right]\,\, ,
\label{kirch03}
\eeqa
where
\beq
\f_{\nu_e}^{\rm eq}=[e^{(\varepsilon_{\nu_e}-(\mu_e-\hat{\mu}))\beta}+1]^{-1}
\label{equileq}
\eeq
is an equilibrium distribution function for the $\nu_e$ neutrino
and it has been assumed that only the electron, proton, and neutron
are in thermal equilibrium.  $\hat{\mu}$ is the difference between the neutron
and the proton chemical potentials.  Note that in $\f_{\nu_e}^{\rm eq}$
there is no explicit reference to a neutrino chemical potential, though
of course in beta equilibrium it is equal to $\mu_e-\hat{\mu}$.
There is no need to construct or
refer to a neutrino chemical potential in neutrino transfer.

We see that eq. (\ref{kirch03}) naturally leads to:
\beq
{\cal {J}}_{net}=\frac{\kappa_a}{1-\f^{\rm eq}_\nu}\left(B_\nu-I_\nu\right)=\kappa_a^*(B_\nu-I_\nu)\,\, .
\label{kirch}
\eeq
Of course, $B_\nu$ is the black body function for neutrinos.
This expression emphasizes the fact that ${\cal C}_{\nu_e n\leftrightarrow e^- p}$ and ${\cal {J}}_{net}$ are the same entity.
If neutrinos were bosons, we would have found a (${1+\f^{\rm eq}_\nu}$) in the denominator, but the
form of eq. (\ref{kirch}) in which $I_\nu$ is manifestly driven to $B_\nu$, the equilibrium
intensity, would have been retained.  From eqs. (\ref{kirch03}) and (\ref{kirch}), we
see that the stimulated absorption correction to $\kappa_a$ is $1/(1-\f^{\rm eq}_\nu)$.
By writing the collision term in the form of eq. (\ref{kirch}), with $\kappa_a$ corrected for stimulated
absorption, we have a net source term that clearly drives $I_\nu$ to equilibrium.  The timescale
is $1/c\kappa_a^*\,$.   Though the derivation of the stimulated absorption correction
we have provided here is for the $\nu_en\leftrightarrow e^-p$ process, this correction is quite general and
applies to all neutrino absorption opacities.

Kirchhoff's Law, expressing detailed balance, is:
\beq
\kappa_a = \eta_\nu/B_\nu \,\,{\rm or}\,\, \kappa^*_a = \eta^{\prime}_\nu/B_\nu\, ,
\label{kirchh}
\eeq
where $\eta^{\prime}_\nu$ is not corrected for final--state neutrino blocking.
Furthermore, the net emissivity can be written as the sum of its {\it spontaneous} and {\it
induced} components:
\beq
\eta_\nu=\kappa_a\left[\frac{B_\nu}{1\pm\f^{\rm eq}_\nu}+\left(1-\frac{1}{1\pm\f^{\rm eq}_\nu}\right)I_\nu\right]\,\,  ,
\label{emission}
\eeq
where $+$ or $-$ is used for bosons or fermions, respectively.
Eq. (\ref{kirchh}) can be used to convert the absorption cross sections described in \S\ref{cross6}
into source terms.

\section{Neutrino Cross Sections}
\label{cross6}

Neutrino--matter cross sections, both for scattering and for absorption,
play the central role in neutrino transport.  The major processes are
the super--allowed charged--current absorptions of $\nu_e$ and $\bar{\nu}_e$
neutrinos on free nucleons, neutral--current scattering off of free nucleons,
alpha particles, and nuclei (Freedman 1974), neutrino--electron/positron scattering,
neutrino--nucleus absorption, neutrino--neutrino scattering, neutrino--antineutrino absorption, and the inverses
of various neutrino production processes such as nucleon--nucleon bremsstrahlung
and the modified URCA process ($\nu_e + n + n \rightarrow e^- + p + n$).
Compared with photon--matter interactions, neutrino-- matter interactions
are relatively simple functions of incident neutrino energy.
Resonances play little or no role and continuum processes dominate. Nice summaries
of the various neutrino cross sections of relevance in supernova
theory are given in \cite{tubbs_schramm} and in \cite{bruenn_1985}.
In particular, \cite{bruenn_1985} discusses in detail
neutrino--electron scattering and neutrino--antineutrino processes
using the full energy redistribution formalism.  He also provides a serviceable approximation
to the neutrino--nucleus absorption cross section (Fuller 1980; Fuller, Fowler, \& Newman 1982; 
Aufderheide et al.~1994). 
For a neutrino energy of $\sim$10 MeV the ratio of the charged--current cross section
to the $\nu_e$--electron scattering cross section is $\sim$100.
However, neutrino--electron scattering does play a role, along with neutrino--nucleon
scattering and nucleon--nucleon bremsstrahlung, in the energy 
equilibration of emergent $\nu_\mu$ neutrinos (Thompson, Burrows, \& Horvath 2000).

Below, we list and discuss many of the absorption and elastic scattering 
cross sections one needs in detailed supernova calculations.
In \S\ref{section:inelastic} and \S\ref{dynamsf}, we provide some straightforward
formulae that can be used to properly handle inelastic scattering.
The set of these processes comprises the essential microphysical 
package for the simulation of neutrino atmospheres and core--collapse supernovae.

\subsection{{\bf $\enu\,+\,n\, \righta\, e^-\,+\,p$:}}
\label{CCabs}
The cross section per baryon for $\nu_e$ neutrino absorption on free neutrons is larger than that
for any other process.   Given the large abundance of free neutrons in protoneutron star atmospheres,
this process is central to $\nu_e$ neutrino transport.
A convenient reference neutrino cross section is $\sigma_o$, given by
\beq
\sigma_o\,=\,\frac{4G^2(m_ec^2)^2}{\pi(\hbar c)^4}\simeq \,1.705\times
10^{-44}\,cm^2\,\, .
\eeq
The total $\enu- n$ absorption cross section is then given by
\beq
\sigma^a_{\enu
n}=\,\sig_o\left(\frac{1+3g_A^2}{4}\right)\,\left(\frac{\varepsilon_{\nu_e}+\Delta_{np}}{m_ec^2}\right)^2\,
\Bigl[1-\left(\frac{m_ec^2}{\varepsilon_{\nu_e}+\Delta_{np}}\right)^2\Bigr]^{1/2}W_M\,\, ,
\label{ncapture}
\eeq
where $g_A$ is the axial--vector coupling constant ($\sim -1.26$), $\Delta_{np}=m_nc^2-m_pc^2=1.29332$
MeV, and for a collision in which the electron gets all of the
kinetic energy $\vep_{e^-}=\varepsilon_{\nu_e}+\Delta_{np}$.
$W_M$ is the correction for weak magnetism and recoil (Vogel 1984) and is approximately equal to $(1 + 1.1\varepsilon_{\nu_e}/m_nc^2)$.
At $\varepsilon_{\nu_e} = 20$ MeV, this correction is only $\sim2.5$\%.  We include it here for
symmetry's sake, since the corresponding correction ($W_{\bar{M}}$) for $\bar{\nu}_e$ neutrino absorption
on protons is $(1 - 7.1\varepsilon_{\bar{\nu}_e}/m_nc^2)$, which at 20 MeV is a large $-15$\%.
To calculate $\kappa_a^*$, $\sigma^a_{\enu n}$ must be multiplied by the stimulated absorption
correction, $1/(1-\f_{\nu_e}\pr)$, and final--state blocking
by the electrons and the protons \`a la eq. (\ref{kirch03}) must be included.

\subsection{{\bf $\aenu\,+\,p\, \righta\, e^+\,+\,n$:}}

The total $\aenu- p$ absorption cross section is given by
\beq
\sigma^a_{\aenu
p}=\sig_o\left(\frac{1+3g_A^2}{4}\right)\,\left(\frac{\vep_{\bar{\nu}_e}-\Delta_{np}}{m_ec^2}\right)^2\,
\Bigl[1-\left(\frac{m_ec^2}{\vep_{\bar{\nu}_e}-\Delta_{np}}\right)^2\Bigr]^{1/2}W_{\bar{M}}\, ,
\label{pcapture}
\eeq
where $\vep_{e^+}=\vep_{\bar{\nu_e}}-\Delta_{np}$ and $W_{\bar{M}}$ is the weak
magnetism/recoil correction given in \S\ref{CCabs}.  Note that $W_{\bar{M}}$ 
is as large as many other corrections and should not be ignored.  To calculate
$\kappa_a^*$, $\sigma^a_{\aenu p}$ must also be corrected for stimulated absorption and final--state blocking.
However, the sign of $\mu_e -\hat{\mu}$ in the stimulated absorption correction for $\bar{\nu}_e$ neutrinos
is flipped, as is the sign of $\mu_e$ in the positron blocking term.  Hence, as a consequence of the severe electron
lepton asymmetry in core--collapse supernovae, both coefficients are very close to one.
Note that the $\aenu\,+\,p\, \righta\, e^+\,+\,n$ process
dominates the supernova neutrino signal in proton--rich underground neutrino
telescopes on Earth, such as Super Kamiokande, LVD, and MACRO, a fact that emphasizes the
interesting complementarities between emission at the supernova and detection in \v Cerenkov and scintillation
facilities.

\subsection{$\enu A \leftrightarrow A\pr e^-$}

>From \cite{bruenn_1985} the total $\nu_e - A$ absorption cross section,
is approximated by
\begin{equation}
\sigma^a_A=\frac{\sigma_o}{14}\,
g_A^2\,N_p(Z)\,N_n(N)\,\left(\frac{\varepsilon_\nu+Q\pr}{m_ec^2}\right)^2\,
\left[1-\left(\frac{m_e c^2}{\varepsilon_\nu+Q\pr}\right)^2\right]^{1/2}
W_{block}
\,\, ,
\end{equation}
where $W_{block} = \left(1-f_{e^-}\right)\,e^{(\mu_n-\mu_p-Q\pr)\beta}$,
$Q\pr=M_{A\pr}-M_A+\Delta\sim\mu_n-\mu_p+\Delta$, $\Delta$ is
the energy of the neutron $1f_{5/2}$ state above the ground state
and is taken to be 3 MeV (Fuller 1982), and
the quantities $N_p(Z)$ and $N_n(N)$ are approximated by:
$N_p(Z)=0$, $Z-20$, and 8 for $Z<20$, $20<Z<28$, and $Z>28$,
respectively. $N_n(N)=6$, $40-N$, and 0 for $N<34$, $43<N<40$, and $N>40$, respectively.
The opacity, corrected for stimulated absorption, is then
\begin{equation}
\kappa_a^*=X_H \,\rho N_A\sigma^a_A(1-\f_{\nu_e}^{\rm eq})^{-1}.
\end{equation}
Since $N_n(N)=0$ for $N>40$, this absorption and emission
process plays a role only during the very early phase of collapse.
Typically at densities near $\rho\sim10^{12}$ g cm$^{-3}$ $\kappa_a^*\rightarrow 0$.

\subsection{{\bf $\nu_i\,+\,p\,\righta\,\nu_i\,+\,p$:}} 
\label{nupro}

The total $\nu_i - p$ scattering cross section for all neutrino species is:
\beq
\sigma_{p}=\frac{\sigma_o}{4}\left(\frac{\epnu}{m_ec^2}\right)^2
\left(4\sin^4\theta_W-2\sinw+\frac{(1+3g_A^2)}{4}\right)\,\, ,
\label{pscatter}
\eeq
where $\theta_W$ is the Weinberg angle and $\sinw \simeq 0.23$.
In terms of $C_V\pr=1/2+2\sinw$ and $C_A\pr=1/2$, eq. (\ref{pscatter}) becomes (Schinder 1990):
\beq
\sigma_{p}=\frac{\sigma_o}{4}\left(\frac{\epnu}{m_ec^2}\right)^2
\left[(C_V\pr-1)^2+3g_A^2(C_A\pr-1)^2\right].
\eeq
The differential cross section is:
\beq
\frac{d\sig_p}{d\Omega}=\frac{\sigma_{p}}{4\pi}(1+\delta_p\mu)\,\, ,
\label{nupscatt}
\eeq
where
\beq
\delta_p=\frac{(C_V\pr-1)^2-g_A^2(C_A\pr-1)^2}{(C_V\pr-1)^2+3g_A^2(C_A\pr-1)^2}\,\, .
\eeq
Note that $\delta_p$, and $\delta_n$ below, are negative ($\delta_p\sim-0.2$ and $\delta_n\sim-0.1$)
and, hence, that these processes are backward--peaked.

The transport (or momentum-transfer) cross section is simply
\beq
\sig^{tr}_p=\frac{\sigma_o}{6}\left(\frac{\epnu}{m_ec^2}\right)^2
\left[(C_V\pr-1)^2+5g_A^2(C_A\pr-1)^2\right]\,\, .
\eeq
where
\begin{equation}
\sigma^{tr}_i=\int\frac{d\sigma_i}{d\Omega}(1-\mu)\,d\Omega
=\sigma_i\left(1-\frac{1}{3}\delta_i\right)\,\,.
\label{transcs}
\end{equation}

\subsection{{\bf $\nu_i\,+\,n\,\righta\,\nu_i\,+\,n$:}} 
\label{nuneu}

The total $\nu_i - n$ scattering cross section is:
\beq
\sigma_{n}=\frac{\sig_o}{4}\left(\frac{\epnu}{m_ec^2}\right)^2
\left(\frac{1+3g_A^2}{4}\right)\,\, .
\label{nunscatt}
\eeq
The corresponding differential cross section is:
\beq
\frac{d\sig_n}{d\Omega}=\frac{\sigma_{n}}{4\pi}(1+\delta_n\mu)\,\, ,
\label{nundiffscatt}
\eeq
where
\beq
\delta_n=\frac{1-g_A^2}{1+3g_A^2}\,\, .
\eeq
The transport cross section is
\beq
\sig^{tr}_n=\frac{\sigma_o}{4}\left(\frac{\epnu}{m_ec^2}\right)^2
\left(\frac{1+5g_A^2}{6}\right)\,\, .
\label{nuntransscatt}
\eeq
The fact that $\delta_p$ and $\delta_n$ are negative and, as a consequence, that
$\sig^{tr}_i$ is greater than $\sigma_i$ increases the neutrino--matter energy coupling rate
for a given neutrino flux in the semi--transparent region.  

\cite{horowitz02} has recently derived expressions that include a weak magnetism/recoil
correction analogous to those previously discussed for the charged-current absorption
rates $\nu_e n\leftrightarrow p e^-$ and $\bar{\nu}_e n\leftrightarrow n e^+$.
We take the following form for the weak magnetism/recoil correction,
a fit to the actual correction factor for the transport cross sections:
\begin{equation}
\sigma^{tr}_{n,p}\rightarrow\sigma^{tr}_{n,p}(1+C_{W_M}\varepsilon_\nu/m_{n,p}),
\end{equation}
where, for neutrino-neutron scattering $C_{W_M}\simeq-0.766$, for neutrino-proton scattering
$C_{W_M}\simeq-1.524$, for anti-neutrino-neutron scattering $C_{W_M}\simeq-7.3656$, and
for anti-neutrino-proton scattering $C_{W_M}\simeq-6.874$.

In fact, neutrino-nucleon scattering is slightly inelastic and when this is germane,
as with mu and tau neutrinos, the more general formalisms of \S\ref{nunucleon} and \S\ref{dynamsf}
are necessary.

\subsection{{\bf $\nu_i\,+\,A\,\righta\,\nu_i\,+\,A$:}}

In the post--bounce phase, nuclei exist in the unshocked region exterior to the shock.
At the high entropies in shocked protoneutron star atmospheres
there are very few nuclei.  There are alpha particles, but their fractional abundances are generally low, growing to
interesting levels due to reassociation of free nucleons just interior to the shock only at late times.
However, nuclei predominate on infall and neutrino-nucleus scattering (Freedman 1974)
is the most important process during the lepton trapping phase.

The differential $\nu_i - A$ neutral--current scattering cross section may be expressed as:
\beq
\frac{d\sig_{A}}{d\Omega}=\frac{\sig_o}{64\pi}
\left(\frac{\varepsilon_\nu}{m_ec^2}\right)^2\,A^2\,
\left\{{\cal W}\,{\cal C}_{FF}+{\cal
C}_{LOS}\right\}^2\,\langle {\cal S}_{ion}\rangle\,(1+\mu)\,\, ,
\label{nunucleuscs}
\eeq
where
\beq
{\cal W} = 1-\frac{2Z}{A}(1-2\sinw) \, ,
\eeq
$Z$ is the atomic number, $A$ is the atomic weight, and $\langle {\cal S}_{ion}\rangle$ is the
ion--ion correlation function, determined mostly by the Coulomb interaction between the
nuclei during infall.
$\langle {\cal S}_{ion}\rangle$,
in eq. (\ref{nunucleuscs}) was investigated by
\cite{horowitz97} who approximated it with the expansion
\begin{equation}
\langle{\cal{S}}_{ion}(\epsilon)\rangle=
\left[1+exp\left(-\sum_{i=0}^6\beta_i(\Gamma)\epsilon^i\right)\right]^{-1}\,\, ,
\end{equation}
where
\begin{equation}
\Gamma=\frac{(Ze)^2}{a}\frac{1}{kT} \hspace{.75cm},\hspace{.75cm}
\epsilon_i=\frac{\varepsilon_{\nu_i}}{\hbar c a}\hspace{.75cm},\hspace{.75cm}
a=\left(\frac{3}{4\pi n_{ion}}\right)^{1/3},
\end{equation}
$a$ is the interparticle spacing, $n_{ion}$ is the number density of ions, $\Gamma$
is the ratio of Coulomb potential between ions to the thermal energy in the medium,
and $\beta_i$ are specified functions of $\Gamma$ for each neutrino species.

\cite{los88} have investigated the electron polarization correction, ${\cal C}_{LOS}$, and
find that
\beq
{\cal C}_{LOS}=
\frac{Z}{A}\left(\frac{1+4\sin^2\theta_W}{1+(kr_D)^2}\right)\,\, ,
\eeq
where the Debye radius is
\beq
r_D=\sqrt{\frac{\pi\hbar^2 c}{4\alpha p_F E_F}} \,\, ,
\eeq
$k^2=|{\bf p-p\pr}|^2=2(\varepsilon_\nu/c)^2(1-\mu)$, $p_F$ and $E_F$ are the electron Fermi momentum
and energy,  and $\alpha$ is the fine--structure constant ($\simeq 137^{-1}$).  Note that
$r_D\sim 10\hbar/p_F$ in the ultra--relativistic limit ($p_F>>m_e c$).
The ${\cal C}_{LOS}$ term is important only for low neutrino energies, generally below $\sim5$ MeV.

Following \cite{tubbs_schramm} and \cite{bml81}, the form factor term, ${\cal
C}_{FF}$, in eq. (\ref{nunucleuscs}) can be approximated by:
\beq
{\cal C}_{FF}=e^{-y(1-\mu)/2}\,\, ,
\eeq
where
$$y=\frac{2}{3}\varepsilon_\nu^2\langle r^2 \rangle/(\hbar c)^2\simeq\left(\frac{\epnu}{56\,
{\rm{MeV}}}\right)^2\left(\frac{A}{100}\right)^{2/3}\, , $$
and $\langle r^2 \rangle^{1/2}$ is the {\it rms} radius of the nucleus.
${\cal C}_{FF}$ differs from $1$ for large $A$ and $\varepsilon_\nu$, when the de Broglie
wavelength of the neutrino is smaller than the nuclear radius.

When $\langle {\cal S}_{ion}\rangle={\cal C}_{FF}={\cal C}_{LOS}+1=1$,
we have simple coherent Freedman scattering.   The physics of the polarization,
ion--ion correlation, and form factor corrections to coherent scattering
is interesting in its own right, but has little effect on supernovae (Bruenn \& Mezzacappa 1997).
The total and transport scattering cross sections for $\nu_i - \alpha$ scattering ($Z=2$;$A=4$) are simply
\begin{equation}
\sigma_\alpha=\frac{3}{2}\sigma^{tr}_\alpha=4\,\sigma_o\left(\frac{\epnu}{m_ec^2}\right)^2\sin^4\theta_W.
\end{equation}

\section{Inelastic Neutrino Scattering}
\label{section:inelastic}

Many authors have studied inelastic neutrino-electron scattering as
an important energy redistribution process which helps to thermalize neutrinos and
increase their energetic coupling to matter in supernova explosions 
(Bruenn 1985; Mezzacappa \& Bruenn 1993abc).
Comparatively little attention has been paid to inelastic
neutrino-nucleon scattering.  \cite{thompson} and \cite{raffelt}
showed that, at least for mu and tau neutrinos, this process cannot 
be ignored.  Here, we review the Legendre expansion formalism
for approximating the angular dependence of the scattering kernel, detail our own
implementation of scattering terms in the Boltzmann equation, and
include a discussion of neutrino-nucleon energy
redistribution. In \S\ref{dynamsf}, we present an alternate approach 
involving dynamical structure factors that is more easily generalized
to include many-body effects.

The general collision integral for inelastic scattering may be written as
\begin{equation}
{\cal L}^{\rm scatt}_\nu[f_\nu]=(1-f_\nu)\int\frac{d^3p_\nu\pr}{c(2\pi\hbar c)^3}f_\nu\pr\,
R^{\rm in}(\varepsilon_\nu,\varepsilon_\nu\pr,\cos\theta) \nonumber
\end{equation}
\begin{equation}
\hspace*{4.8cm}-f_\nu\int\frac{d^3p_\nu\pr}{c(2\pi\hbar c)^3}(1-f_\nu\pr)\,
R^{\rm out}(\varepsilon_\nu,\varepsilon_\nu\pr,\cos\theta)
\label{gencoll}
\end{equation}
\begin{equation}
=\tilde{\eta}_\nu^{\rm scatt}-\tilde{\chi}_\nu^{\rm scatt}f_\nu\hspace{2.5cm}
\end{equation}
where $\cos\theta$ is the cosine of the scattering angle, $\varepsilon_\nu$ is the
incident neutrino energy,
and $\varepsilon_\nu\pr$ is the scattered neutrino energy.   Although we suppress it here, the incident
and scattered neutrino phase space distribution functions ($f_\nu$ and $f_\nu\pr$, respectively)
have the following dependencies: $f_\nu=f_\nu(r,t,\mu,\varepsilon_\nu)$ and
$f_\nu\pr=f_\nu\pr(r,t,\mu\pr,\varepsilon_\nu\pr)$.  
$\mu$ and $\mu\pr$ are the cosines of the angular
coordinate of the zenith angle in spherical symmetry and are related to $\cos\theta$ through
\begin{equation}
\cos\theta=\mu\mu\pr+[(1-\mu^2)(1-\mu^{\prime\,2})]^{1/2}\cos(\phi-\phi\pr).
\label{costheta}
\end{equation}
The only difference
between $f_\nu$ and $\f_{\nu}$ in \S\ref{stimabs} is that here $f_\nu$ has explicit $\mu$
and $\varepsilon_\nu$ dependencies.
$R^{\rm in}$ is the scattering kernel for scattering into the
bin ($\varepsilon_\nu$, $\mu$) from any bin ($\varepsilon_\nu\pr$, $\mu\pr$)
and $R^{\rm out}$ is the scattering kernel for scattering out of the
bin ($\varepsilon_\nu$, $\mu$) to any bin ($\varepsilon_\nu\pr$, $\mu\pr$).
The kernels are Green functions that connect points
in energy and momentum space.  One may also write
$R(\varepsilon_\nu,\varepsilon_\nu\pr,\cos\theta)$ as $R(q,\omega)$, where
$\omega(=\varepsilon_\nu-\varepsilon_\nu\pr)$ is the energy transfer and
$q(=[\varepsilon_\nu^2+\varepsilon_\nu^{\prime\,2}-2\varepsilon_\nu\varepsilon_\nu\pr\cos\theta]^{1/2})$
is the momentum transfer, so that the kernel explicitly reflects these dependencies (\S\ref{dynamsf}).

An important simplification comes from detailed balance, a consequence
of the fact that these scattering rates must drive the distribution to equilibrium.
One obtains: $R^{\rm in}=e^{-\beta\omega}R^{\rm out}$, where $\beta=1/T$.  Therefore, we need deal
only with $R^{\rm out}$.  The scattering kernels for 
inelastic neutrino-nucleon and neutrino-electron scattering
depend in a complicated fashion on scattering angle.  
For this reason, one generally approximates the angular dependence
of the scattering kernel with a truncated Legendre series (Bruenn 1985).  We take
\begin{equation}
R^{\rm out}(\varepsilon_\nu,\varepsilon_\nu\pr,\cos\theta)
=\sum_{l=0}^\infty\frac{2l+1}{2}\Phi(\varepsilon_\nu,\varepsilon_\nu\pr,\cos\theta)
P_l(\cos\theta),
\end{equation}
where
\begin{equation}
\Phi_l(\varepsilon_\nu,\varepsilon_\nu^\prime)=\int_{-1}^{+1}d(\cos\theta)\,
R^{\rm out}(\varepsilon_\nu,\varepsilon_\nu^\prime,\cos\theta)P_l(\cos\theta).
\label{momentkernel}
\end{equation}
In practice, one expands only to first order so that
\begin{equation}
R^{\rm out}(\varepsilon_\nu,\varepsilon_\nu\pr,\cos\theta)\sim
\frac{1}{2}\Phi_0(\varepsilon_\nu,\varepsilon_\nu\pr)+
\frac{3}{2}\Phi_1(\varepsilon_\nu,\varepsilon_\nu\pr)\cos\theta.
\label{kernelapprox}
\end{equation}
Substituting into the first term on the right-hand-side of eq.~(\ref{gencoll})
(the source) gives
\begin{equation}
\tilde{\eta}_\nu^{\rm scatt}=(1-f_\nu)
\int_0^\infty \frac{d\varepsilon_\nu\pr\varepsilon_\nu^{\prime\,2}}{c(2\pi\hbar c)^3}\,e^{-\beta\omega}
\int_{-1}^{+1}d\mu\pr f_\nu\pr\int_0^{2\pi}d\phi\pr
\left[\frac{1}{2}\Phi_0+
\frac{3}{2}\Phi_1\cos\theta\right]
\end{equation}
Substituting for $\cos\theta$ using eq.~(\ref{costheta}) and using the definitions
\begin{equation}
\tilde{J}_\nu=\frac{1}{2}\int_{-1}^{+1}d\mu f_\nu
\label{jtilde}
\end{equation}
and
\begin{equation}
\tilde{H}_\nu=\frac{1}{2}\int_{-1}^{+1}d\mu \mu f_\nu
\label{htilde}
\end{equation}
we have that
\begin{equation}
\tilde{\eta}_\nu^{\rm scatt}=(1-f_\nu)\frac{4\pi}{c(2\pi\hbar c)^3}
\int_0^\infty d\varepsilon_\nu\pr \varepsilon_\nu^{\prime\,2} e^{-\beta\omega}
\left[\frac{1}{2}\Phi_0\tilde{J}_\nu\pr+\frac{3}{2}\Phi_1\mu\tilde{H}_\nu\pr\right].
\end{equation}
Integrating over $\mu$ to get the source for the zeroth moment of the transport equation,
\begin{equation}
\frac{1}{2}\int_{-1}^{+1}d\mu\,\tilde{\eta}_\nu^{\rm scatt}=
\frac{4\pi}{c(2\pi\hbar c)^3}
\int_0^\infty d\varepsilon_\nu\pr \varepsilon_\nu^{\prime\,2} e^{-\beta\omega}
\left[\frac{1}{2}\Phi_0\tilde{J}_\nu\pr(1-\tilde{J}_\nu)-\frac{3}{2}\Phi_1\tilde{H}_\nu\tilde{H}_\nu\pr\right].
\label{jtildeeta}
\end{equation}
Similarly, we may write the sink term of the Boltzmann equation  collision term 
(second term in eq.~\ref{gencoll}), employing the Legendre expansion
\begin{equation}
\tilde{\chi}_\nu^{\rm scatt}=\frac{4\pi}{c(2\pi\hbar c)^3}
\int_0^\infty d\varepsilon_\nu\pr \varepsilon_\nu^{\prime\,2}
\left[\frac{1}{2}\Phi_0(1-\tilde{J}_\nu\pr)-\frac{3}{2}\Phi_1\mu\tilde{H}_\nu\pr\right].
\end{equation}
The contribution to the zeroth moment equation is then
\begin{equation}
\frac{1}{2}\int_{-1}^{+1}d\mu(-\tilde{\chi}_\nu^{\rm scatt}f_\nu)=-\frac{4\pi}{c(2\pi\hbar c)^3}
\int_0^\infty d\varepsilon_\nu\pr \varepsilon_\nu^{\prime\,2}
\left[\frac{1}{2}\Phi_0(1-\tilde{J}_\nu\pr)\tilde{J}_\nu-\frac{3}{2}\Phi_1\tilde{H}_\nu\tilde{H}_\nu\pr\right].
\label{jtildechi}
\end{equation}
Combining these equations, we find that
$$\frac{1}{2}\int_{-1}^{+1}d\mu\,{\cal L}^{\rm scatt}_\nu[f_\nu]=
\frac{4\pi}{c(2\pi\hbar c)^3}\int_0^\infty d\varepsilon_\nu\pr \varepsilon_\nu^{\prime\,2} \hspace{6cm}$$
\begin{equation}
\hspace*{1cm}\,\,\,\times\,\,\,
\left\{\frac{1}{2}\Phi_0\left[\tilde{J}_\nu\pr(1-\tilde{J}_\nu)e^{-\beta\omega}-(1-\tilde{J}_\nu\pr)\tilde{J}_\nu \right]
-\frac{3}{2}\Phi_1\tilde{H}_\nu\tilde{H}_\nu\pr(e^{-\beta\omega}-1)\right\}.
\end{equation}
One can see immediately that including another term in the Legendre expansion
(taking $R^{\rm out}\sim(1/2)\Phi_0+(3/2)\Phi_1\cos\theta+
(5/2)\Phi_2(1/2)(3\cos^2\theta-1)$) necessitates including $\tilde{P}_\nu$ and
$\tilde{P}_\nu\pr$, the second angular moment of the neutrino phase-space distribution
function, in the source and sink terms.  
While easily doable, we advocate retaining only
the linear term and explore this approximation in the next two subsections.

\subsection{Neutrino-Electron Scattering}
\label{app:escatt}

The opacity due to neutrino-electron scattering can be large compared with that of
other processes at low neutrino energies ($\varepsilon_\nu\lesssim5$ MeV)
and at high matter temperatures.
A good approximation to the total scattering cross section has been derived by \cite{bowers},
which interpolates between analytic limits derived in \cite{tubbs_schramm}:
\begin{equation}
\sigma_e=\frac{3}{8}\sigma_o\,(m_e c^2)^{-2}\,\,\varepsilon_\nu\left(T+\frac{\mu_e}{4}\right)
\left[(C_V+C_A)^2+\frac{1}{3}(C_V-C_A)^2\right],
\label{approxe}
\end{equation}
where $C_V=1/2+2\sinw$ for electron types, $C_V=-1/2+2\sinw$ for mu and tau neutrino types,
$C_A=+1/2$ for $\nu_e$ and $\aunu$, and $C_A=-1/2$ for $\aenu$ and $\unu$.

However, the use of cross section (\ref{approxe}) implicitly 
ignores the ineleasticity of neutrino-electron
scattering.  When inelasticity is germane and a full energy redistribution
formalism is needed, the scattering kernel approach of \S\ref{section:inelastic} in which
different energy groups are coupled must be employed.
The scattering kernel $R(\varepsilon_\nu,\varepsilon_\nu\pr,\cos\theta)$ in \S\ref{section:inelastic}
is related to the fully relativistic structure function for neutrino-electron scattering:
\begin{equation}
R^{\rm out}(\varepsilon_\nu,\varepsilon_\nu\pr,\cos\theta)=
2G^2\frac{q_\mu^2}{\varepsilon_\nu\pr\varepsilon_\nu}
[A{\cal S}_1(q,\omega)+{\cal S}_2(q,\omega)+B{\cal S}_3(q,\omega)](1-e^{-\beta\omega})^{-1},
\end{equation}
where $q_\mu\left(=(\omega,\vec{q}\,)\right)$ is the four-momentum transfer, 
$A=(4\varepsilon_\nu\varepsilon_\nu\pr+q_\alpha^2)/2q^2$, 
$B=\varepsilon_\nu+\varepsilon_\nu\pr$, and $q_\alpha=(\varepsilon_\nu,\vec{q}_{\nu})$.
The relativistic structure functions (${\cal S}_i$) are written in terms of the imaginary
parts of the retarded polarization functions (Reddy, Prakash, \& Lattimer 1998; 
Thompson, Burrows, \& Horvath 2000; \S\ref{dynamsf}):
\beq
{\cal S}_1(q,\omega)\,=
\,({\cal{V}}^2+{\cal{A}}^2)\,\,\left[\,{\rm Im}\Pi_L^R(q,\omega)+
{\rm Im}\Pi_T^R(q,\omega)\,\right],
\label{s1}
\eeq
\beq
{\cal S}_2(q,\omega)\,=
\,({\cal{V}}^2+{\cal{A}}^2)\,\,{\rm Im}\Pi_T^R(q,\omega)-
{\cal{A}}^2{\rm Im}\Pi_A^R(q,\omega),
\label{s2}
\eeq
and
\beq   
{\cal S}_3(q,\omega)\,=
\,2{\cal{V}}\hspace{-.06cm}{\cal{A}}\,\,{\rm Im}\Pi_{VA}^R(q,\omega).
\label{s3}
\eeq
$\cal{V}$ and $\cal{A}$ are the appropriate vector and
axial-vector coupling constants.
Each of the polarization functions can be written in terms of one-dimensional integrals over
electron energy ($\varepsilon_e$),  which we label $I_n$ (Reddy, Prakash, \& Lattimer 1998)
\begin{equation}
{\rm Im}\Pi_L^R(q,\omega)\,=
\,\frac{q_\mu^2}{2\pi|q|^3}\left[I_2+\omega I_1+
\frac{q_\mu^2}{4}I_0\right],
\end{equation}
\begin{equation}
{\rm Im}\Pi_T^R(q,\omega)\,=
\,\frac{q_\mu^2}{4\pi|q|^3}\left[I_2+\omega I_1+
\left(\frac{q_\mu^2}{4}+\frac{q^2}{2}+
m^2\frac{q^2}{q_\mu^2}\right)I_0\right],
\end{equation}
\begin{equation}
{\rm Im}\Pi_A^R(q,\omega)\,=\,\frac{m^2}{2\pi|q|}I_0,
\end{equation}
and
\begin{equation}
{\rm Im}\Pi_{VA}^R(q,\omega)\,=
\,\frac{q_\mu^2}{8\pi|q|^3}\left[\omega I_0+2I_1\right].
\end{equation}
\cite{reddy_1998} were able to express the
$I_n$'s in terms of polylogarithmic integrals such that
\begin{equation}
I_0\,=\,Tz\left(1-\frac{\xi_1}{z}\right),
\label{I0}
\end{equation}
\begin{equation}
I_1\,=\,T^2z\left(\eta_e-\frac{z}{2}-\frac{\xi_2}{z}-
\frac{e_-\xi_1}{zT}\right),
\label{I1}
\end{equation}
and
\begin{equation}
I_2\,=\,T^3z\left(\eta_e^2-z\eta_e+\frac{\pi^2}{3}+
\frac{z^2}{3}+2\frac{\xi_3}{z}-
2\frac{e_-\xi_2}{Tz}+\frac{e_-^2\xi_1}{T^2z}\right),
\label{I2}
\end{equation}
where $\eta_e=\mu_e/T$ is the electron degeneracy,
$z=\beta\omega$, $\omega$ is the energy transfer, and
\begin{equation}
e_-=-\frac{\omega}{2}+\frac{q}{2}\sqrt{1-4\frac{m^2}{q_\mu^2}}.
\end{equation}
In eqs. (\ref{I0}-\ref{I2}), the $\xi_n$'s are differences
between polylogarithmic integrals;
$\xi_n={\rm Li}_n(-\alpha_1)-{\rm Li}_n(-\alpha_2)$, where
\begin{equation}
{\rm Li}_n(y)=\int_0^y\frac{{\rm Li}_{n-1}(x)}{x}\,dx,
\end{equation}
and Li$_1(x)={\rm ln}(1-x)$.  The arguments necessary for
computing the integrals are
$\alpha_1={\rm exp}[\beta(e_-+\omega)-\eta_e]$ and
$\alpha_2={\rm exp}(\beta e_--\eta_e)$.  Tables for computation of ${\rm Li}_n(y)$
and the $I_n$s can be provided by Sanjay Reddy.

\begin{figure} 
\vspace*{6.0in}
\hbox to\hsize{\hfill\includegraphics{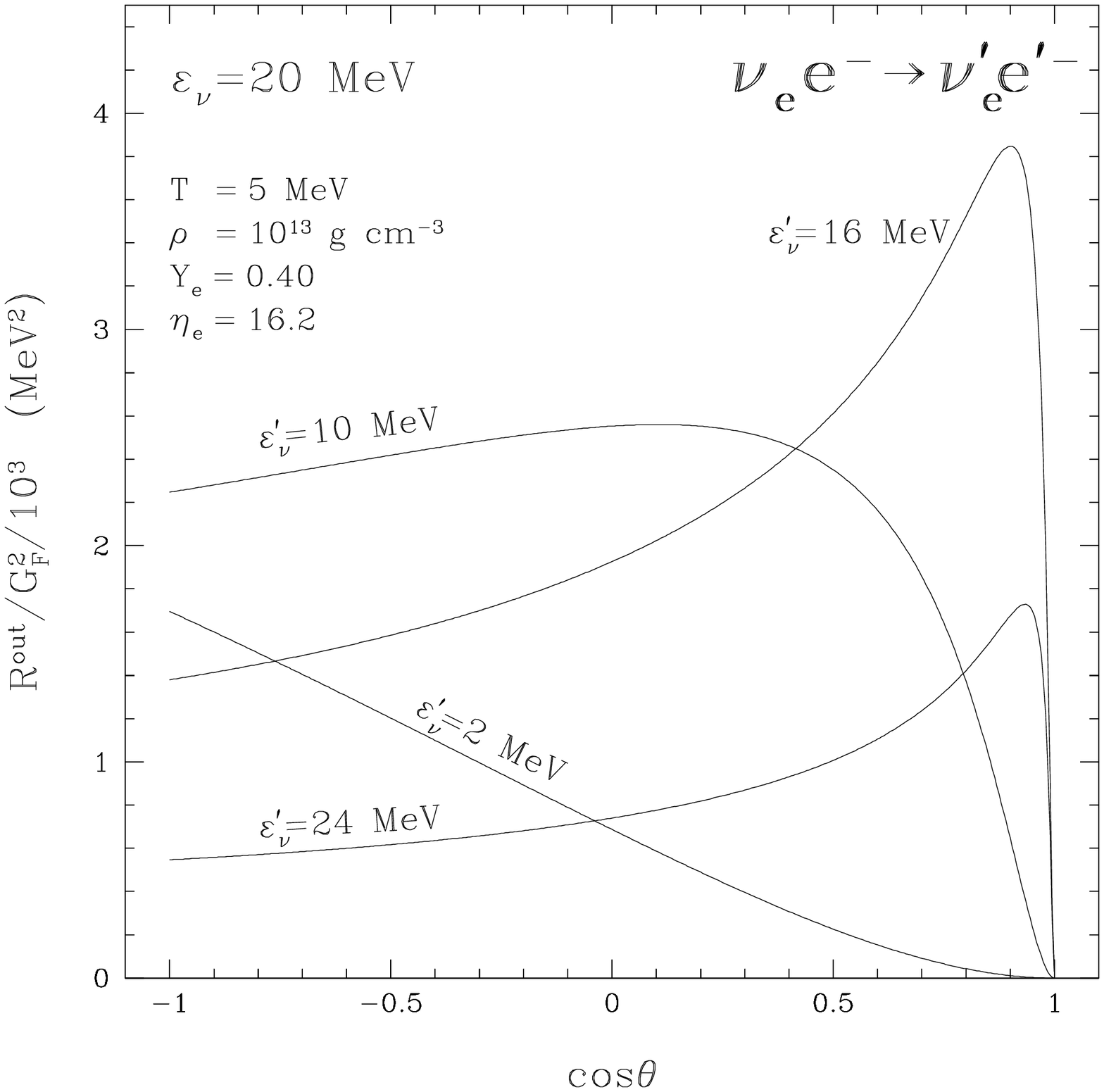}\kern+6in\hfill}
\caption[$R^{\rm out}(\varepsilon_\nu,\varepsilon_\nu\pr,\cos\theta)$ for $\nu_e e^-$
scattering vs.~$\cos\theta$]
{The scattering kernel 
$R^{\rm out}(\varepsilon_\nu,\varepsilon_\nu\pr,\cos\theta)$ for
$\nu_e-$electron scattering as a function of
$\cos\theta$ for $\varepsilon_\nu=20$ MeV and $\varepsilon_\nu\pr=2$, 10, 16, and 24 MeV,
at a representative thermodynamic point ($T=5$ MeV, $\rho=10^{13}$ g cm$^{-3}$, $Y_e=0.4$).
}
\label{ek1}
\end{figure}

\begin{figure} 
\vspace*{6.0in}
\hbox to\hsize{\hfill\includegraphics{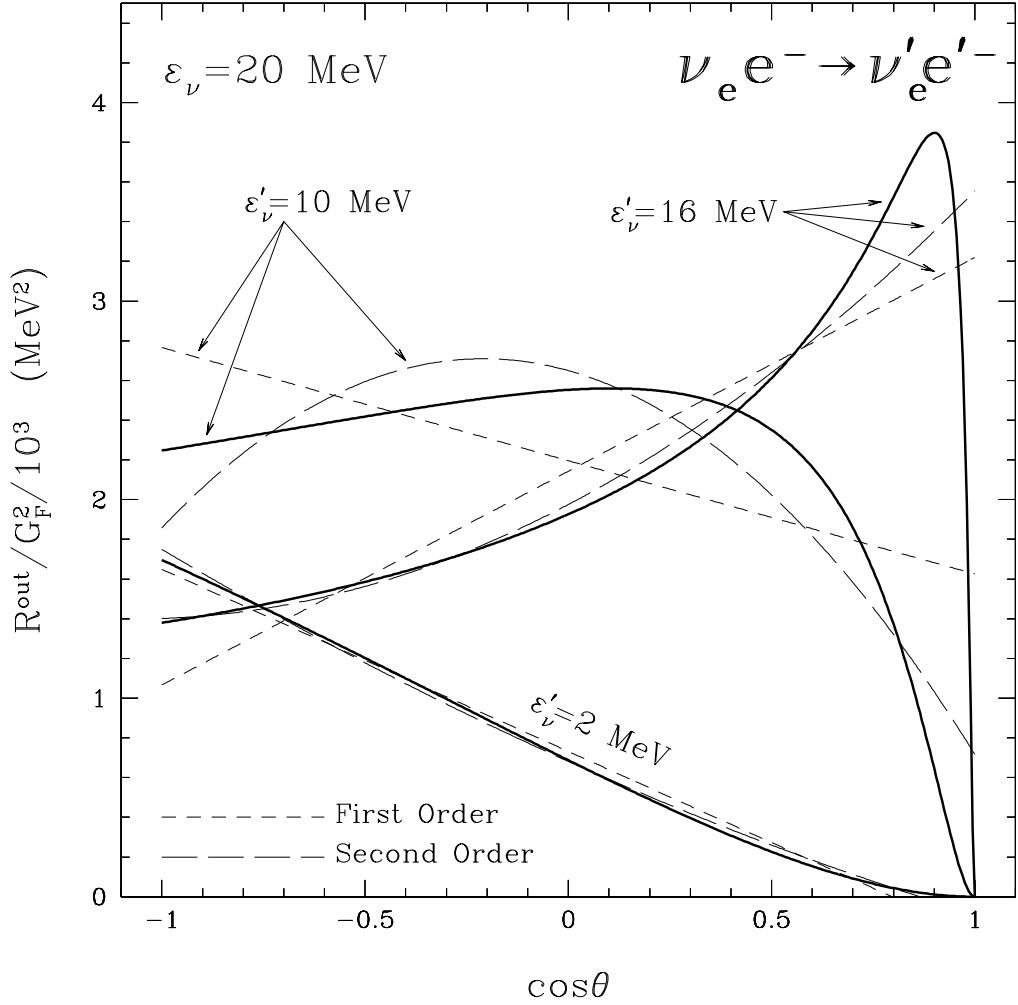}\kern+6in\hfill}
\caption[$R^{\rm out}(\varepsilon_\nu,\varepsilon_\nu\pr,\cos\theta)$ for $\nu_e e^-$
scattering vs.~$\cos\theta$ with 0th, 1st, \& 2nd-order Legendre expansions]
{For the same thermodynamic point as used for Fig.~(\ref{ek1}),
the scattering kernel ($R^{\rm out}$, thick solid lines) for $\nu_e-$electron scattering 
as a function of $\cos\theta$, 
for $\varepsilon_\nu=20$ MeV and $\varepsilon_\nu\pr=2$, 10, and 16 MeV. 
Short dashed lines show the first-order Legendre series expansion approximation
to $R^{\rm out}$, which is linear in $\cos\theta$; $R^{\rm out}\sim(1/2)\Phi_0+(3/2)\Phi_1\cos\theta$.
The long dashed line shows the improvement in going to second order in $\cos\theta$
by taking $R^{\rm out}\sim(1/2)\Phi_0+(3/2)\Phi_1\cos\theta+
(5/2)\Phi_2(1/2)(3\cos^2\theta-1)$.
}
\label{ek2}
\end{figure}

\begin{figure} 
\vspace*{6.0in}
\hbox to\hsize{\hfill\includegraphics{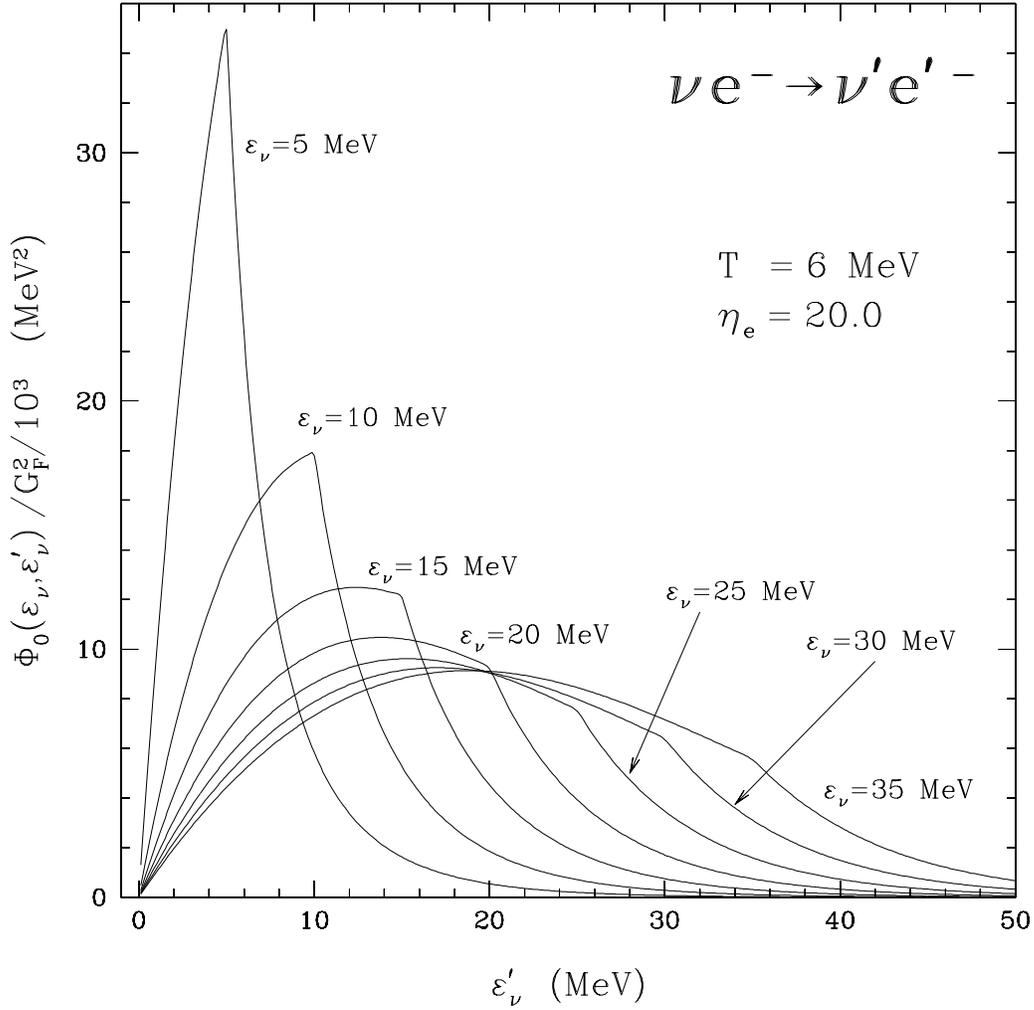}\kern+6in\hfill}
\caption[$\Phi_0(\varepsilon_\nu,\varepsilon_\nu\pr)$ vs.~$\varepsilon_\nu\pr$ 
for $\nu_e e^-$ scattering for various $\varepsilon_\nu$]
{The $l=0$ term in the Legendre expansion of the $\nu_e-$electron scattering
kernel, $\Phi_0(\varepsilon_\nu,\varepsilon_\nu\pr)$ (eq.~\ref{momentkernel}), 
for $T=6$ MeV and $\eta_e=20$ as a 
function of $\varepsilon_\nu\pr$ for $\varepsilon_\nu=5$, 10, 15, 20, 25, and 35 MeV.
Note that for any $\varepsilon_\nu$, the neutrino is predominantly downscattered.
The magnitude of $\Phi_0(\varepsilon_\nu,\varepsilon_\nu\pr)$ and sign of $\langle\omega\rangle$
are to be compared with those in Fig.~(\ref{nk3}).}
\label{ek3}
\end{figure}

Figure (\ref{ek1}) shows the full scattering kernel for $\varepsilon_\nu=20$ MeV
and $\varepsilon_\nu\pr=2$, 10, and 16 MeV as a function of the cosine of the
scattering angle, $\cos\theta$.
Note that although the absolute value of the energy transfer ($|\varepsilon_\nu-\varepsilon_\nu\pr|$)
is the same for both $\varepsilon_\nu\pr=16$ MeV and
$\varepsilon_\nu\pr=24$ MeV, the absolute value of $R^{\rm out}(20,16,\cos\theta)$ is
more than twice that of $R^{\rm out}(20,24,\cos\theta)$, reflecting the fact that
at this temperature the incoming neutrino is more likely to downscatter than upscatter.
Figure (\ref{ek2}) shows the scattering kernel for the same conditions as Fig.~(\ref{ek1}),
but also includes both the first-order (short dashed lines) and second-order (long
dashed lines) approximations to $R^{\rm out}$.  We generally employ the former.  The latter
is included to illustrate the improvement in including higher-order terms.  In fact,
the actual degree of expansion necessary to capture accurately the physics can only be
ascertained by running full transport calculations.  We have run dynamical simulations with only
the zeroth-order and first-order terms in the Legendre expansion and find little or no difference
between the emergent spectra and detailed thermodynamical evolution in the models
we have studied.  \cite{smitthesis} and \cite{smit96} have
explored the importance of including the second-order
term ($\propto\cos^2\theta$, shown here) and find it negligible.
The scattering-angle-averaged kernel, also the zeroth-order term in the Legendre series for
$R^{\rm out}$, is shown in Fig.~(\ref{ek3}) for $\nu_e-$electron scattering
for a matter temperature ($T$) of 6 MeV and with an electron degeneracy factor $\eta_e=\mu_e/T=20$
as a function of $\varepsilon_\nu\pr$ for various incoming neutrino energies, $\varepsilon_\nu$s.

In a full time-dependent simulation it is numerically costly to compute
the Legendre moments of the scattering kernel ($\Phi_0(\varepsilon_\nu,\varepsilon_\nu\pr)$ and
$\Phi_1(\varepsilon_\nu,\varepsilon_\nu\pr)$) via eq.~(\ref{momentkernel}) at
each point in the computational domain.  For this reason we tabulate 
$\Phi_0(\varepsilon_\nu,\varepsilon_\nu\pr)$ and $\Phi_1(\varepsilon_\nu,\varepsilon_\nu\pr)$ 
for each $\varepsilon_\nu$ and $\varepsilon_\nu\pr$ pair on a grid in 
temperature and $\eta_e$.  Because the  vector and axial-vector coupling constants 
for neutrino-electron scattering and the neutrino energy grouping differ between 
$\nu_e$, $\bar{\nu}_e$, and $\nu_\mu$ neutrinos, we construct separate tables for each species.
Gauss-Legendre quadratures (16-point) are used to evaluate the angular integrals
over $\cos\theta$ for $l=0$ and $l=1$ in eq.~(\ref{momentkernel}).
At a given temperature/density/composition point the equation of state returns $\eta_e$
and we then perform a six-point bivariant interpolation in $T$-$\eta_e$ space, for the given
$\varepsilon_\nu$-$\varepsilon_\nu\pr$ combination, to obtain 
$\Phi_0(\varepsilon_\nu,\varepsilon_\nu\pr)$ and $\Phi_1(\varepsilon_\nu,\varepsilon_\nu\pr)$.
The integrals over $\varepsilon_\nu\pr$ for each energy, 
which yield $\tilde{\eta}_\nu^{\rm scatt}$ and $\tilde{\chi}_\nu^{\rm scatt}$,
are then computed using simple trapezoidal rule integration.
In practice, we use 40 energy groups ($n_f$),
30 temperature points ($N_T$), and 30 $\eta_e$ points ($N_\eta$).
The tables are then $l\times n_f\times n_f\pr\times N_T\times N_\eta$
in size, with $l=2$ ($\Phi_0$ and $\Phi_1$), or approximately 50 Megabytes. 
The source and sink at each energy ($\tilde{\eta}_\nu^{\rm scatt}$ and $\tilde{\chi}_\nu^{\rm scatt}$)
are then included {\it explicitly} in a manner analogous to any of the emission 
or absorption processes. 
Using this method, our calculations including neutrino-electron scattering,
are just 10-15\% slower than our calculations ignoring this 
important equilibration process (Thompson, Burrows, \& Pinto 2002).

This calculation of $\tilde{\eta}_\nu^{\rm scatt}$ and $\tilde{\chi}_\nu^{\rm scatt}$
uses only values of the neutrino energy density and flux from the previous timestep
and, hence, we introduce an explicit timescale into the energy and electron fraction 
updates returned by our transport algorithm. For this reason, when the scattering rate is large 
we may encounter a numerical instability.  Because the largest scattering rates are encountered
when the neutrino phase-space distribution function is in local thermodynamical equilibrium
and the scattering off electrons is unimportant, we simply divide the source and sink by
a large factor (typically 100 above $\rho=10^{14}$ g cm$^{-3}$), thus circumventing 
the problem of introducing a short and explicit timescale.  Again, because 
$f_\nu=f_\nu^{\rm eq}$ at these high densities, this approximation is acceptable.

\subsection{Neutrino-Nucleon Scattering}
\label{nunucleon}

The kernel for inelastic neutrino-nucleon scattering can be related to the non-relativistic
structure function employed in the thermalization studies of \cite{thompson}:
\begin{equation}
R^{\rm out}(\varepsilon_\nu,\varepsilon_\nu\pr,\cos\theta)=
G^2 S(q,\omega)[(1+\cos\theta)V^2+(3-\cos\theta)A^2],
\end{equation}
where $S(q,\omega)$ is given in terms of the
imaginary part of the polarization
function, analogous to neutrino-electron scattering (\S\ref{app:escatt}).

\begin{figure} 
\vspace*{6.0in}
\hbox to\hsize{\hfill\includegraphics{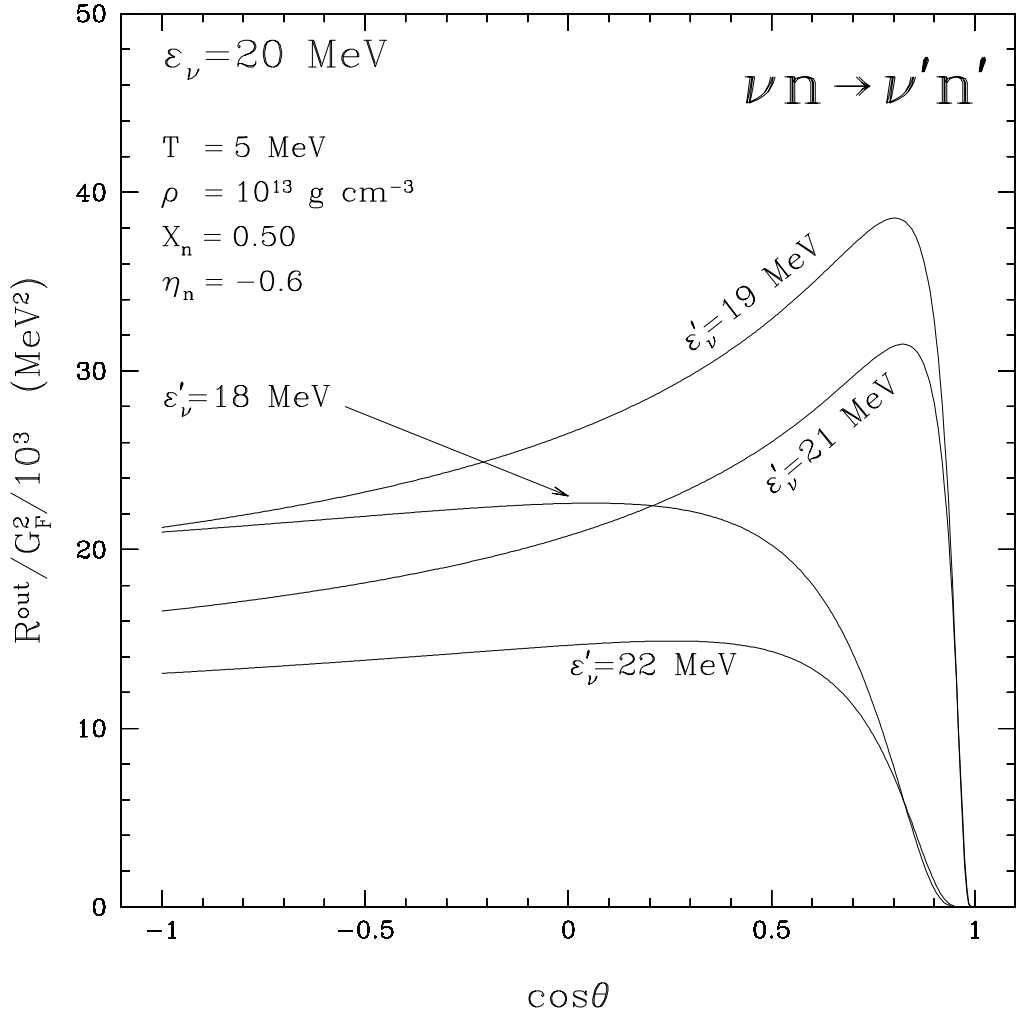}\kern+6in\hfill}
\caption[$R^{\rm out}(\varepsilon_\nu,\varepsilon_\nu\pr,\cos\theta)$ for $\nu_e n$
scattering vs.~$\cos\theta$]
{The scattering kernel 
$R^{\rm out}(\varepsilon_\nu,\varepsilon_\nu\pr,\cos\theta)$ for
$\nu_e-$neutron scattering as a function of
$\cos\theta$ for $\varepsilon_\nu=20$ MeV and $\varepsilon_\nu\pr=18$, 19, 21, and 22 MeV,
at a representative thermodynamic point ($T=5$ MeV, $\rho=10^{13}$ g cm$^{-3}$, $X_n=0.5$).
Note that although the absolute value of the energy transfer 
($|\varepsilon_\nu-\varepsilon_\nu\pr|$) is the same for 
both $\varepsilon_\nu\pr=19$ MeV and 
$\varepsilon_\nu\pr=21$, the absolute value of $R^{\rm out}(20,19,\cos\theta)$ is
greater than that of $R^{\rm out}(20,21,\cos\theta)$, reflecting the fact that
at this temperature the incoming neutrino is more likely to downscatter than upscatter.
}
\label{nk1}
\end{figure}

\begin{figure} 
\vspace*{6.0in}
\hbox to\hsize{\hfill\includegraphics{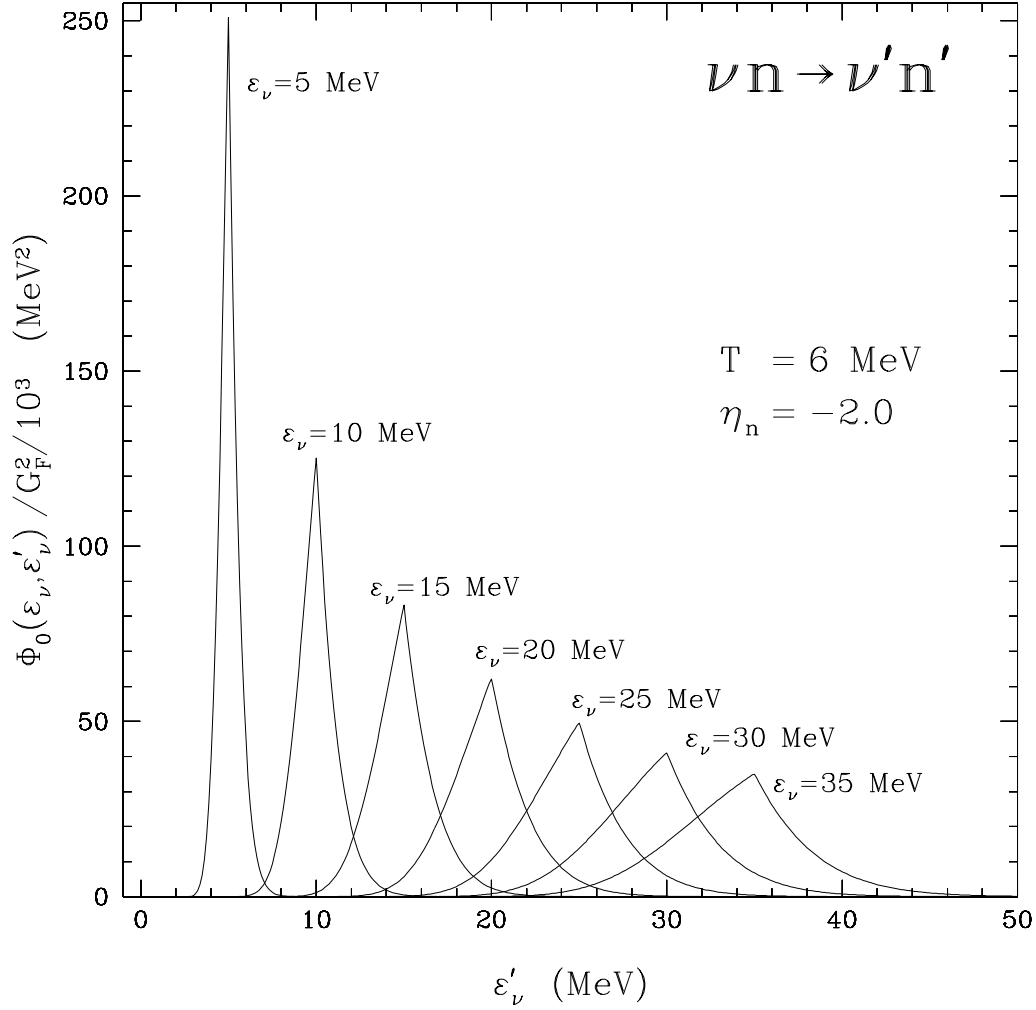}\kern+6in\hfill}
\caption[$\Phi_0(\varepsilon_\nu,\varepsilon_\nu\pr)$ vs.~$\varepsilon_\nu\pr$ 
for $\nu_e n$ scattering for various $\varepsilon_\nu$]
{The $l=0$ term in the Legendre expansion of the neutrino-nucleon scattering
kernel, $\Phi_0(\varepsilon_\nu,\varepsilon_\nu\pr)$ (eq.~\ref{momentkernel}), for 
$T=6$ MeV and $\eta_n=-2$ as a 
function of $\varepsilon_\nu\pr$ for $\varepsilon_\nu=5$, 10, 15, 20, 25, and 35 MeV.
Note that for $\varepsilon_\nu=5$ MeV the neutrino is predominantly upscattered, while
for $\varepsilon_\nu=35$ MeV the neutrino is predominantly downscattered.  The magnitude 
of $\Phi_0(\varepsilon_\nu,\varepsilon_\nu\pr)$ and sign of $\langle\omega\rangle$
are to be compared with those in Fig.~(\ref{ek3}).}
\label{nk3}
\end{figure}

The neutrino-nucleon scattering kernels, while larger in absolute magnitude than the
corresponding neutrino-electron scattering kernels are, at most points in energy space and thermodynamic space,
much less broad.  In fact, in thermodynamic regimes most relevant for the formation of the
various species' spectra, the kernel is quite sharply peaked in energy.  That is, for a given
$\varepsilon_\nu$ the distribution of $\varepsilon_\nu\pr$'s is tightly centered on $\varepsilon_\nu$
because the ratio of the neutrino energy to the nucleon mass is small.
$R^{\rm out}$ as a function of $\cos\theta$, for neutrino-neutron scattering, is shown in Fig.~(\ref{nk1}) at a representative
thermodynamic point. This plot is analogous to Fig.~(\ref{ek1}) for neutrino-electron scattering.
By comparing the $\varepsilon_\nu\pr=19$ MeV
line with that for $\varepsilon_\nu\pr=21$ MeV, one sees that the former is larger and, hence, downscattering
is preferred.  In addition, the overall magnitude is much larger than in the neutrino-electron scattering case.
Figure (\ref{nk3}) shows $\Phi_0$ (eq.~\ref{momentkernel}) for neutrino-neutron scattering as a function
of $\varepsilon_\nu\pr$ for several $\varepsilon_\nu$'s at the same thermodynamic point
used in Fig.~(\ref{ek3}),
the corresponding figure for neutrino-electron scattering.  Note that while
downscatting is strong for the $\varepsilon_\nu=35$ MeV kernel, there 
is almost equal upscattering at $\varepsilon_\nu=5$ MeV.  
In order to explore the effect of this process on the emergent spectra in
dynamical simulations we must first deal with a technical problem.

In a typical simulation, we employ 40 energy groups for all neutrino species with
$1\,{\rm MeV}\leq\varepsilon_\nu\leq320\,{\rm MeV}$ for electron-type neutrinos and
$1\,{\rm MeV}\leq\varepsilon_\nu\leq100\,{\rm MeV}$ for anti-electron and muon neutrinos.
The grouping is generally logarithmic for the $\nu_e$s and linear for $\bar{\nu}_e$ and $\nu_\mu$.
Neutrino-electron scattering and neutrino-nucleon scattering are most important as thermalization
mechanisms at energies below $\sim60$ MeV, where the phase space distribution of all neutrino species
is largest.  One can see clearly from Fig.~(\ref{nk3}) that a trapezoidal rule integration of
$\Phi_0$ over $\varepsilon_\nu\pr$ as it appears in eq.~(\ref{jtildeeta}) and eq.~(\ref{jtildechi})
may grossly overestimate $\tilde{\eta}_\nu^{\rm scatt}$ and $\tilde{\chi}_\nu^{\rm scatt}$.
In fact, with logarithmic energy grouping one may even calculate upscattering when there is
none because the energy groups become larger with increasing energy.

For neutrino-electron scattering, we are able to employ a simple trapezoidal rule and adequately
capture the qualities of the kernel.
This implies that we use a linear interpolation for $\tilde{J}_\nu\pr$
and $\tilde{H}_\nu\pr$ in each energy bin.  As Fig.~(\ref{nk3}) shows, however, in order to get
an accurate integral over $\Phi_0(\varepsilon_\nu,\varepsilon_\nu\pr)$, we must do better than simple
trapezoidal rule with linear interpolation.
For neutrino-nucleon scattering, in order to increase the accuracy of our scheme
without compromising computational efficiency, for a given energy grouping we pre-compute
a grid of integrals over $\varepsilon_\nu\pr$.  We assume that during the dynamical calculation
and the computation of $\tilde{\eta}_\nu^{\rm scatt}$ and $\tilde{\chi}_\nu^{\rm scatt}$,
both $\tilde{J}_\nu\pr$ and $\tilde{H}_\nu\pr$ are proportional to $A\varepsilon_\nu\pr+B$
over an energy interval $\varepsilon_{\nu,\,i}\leq\varepsilon_\nu\pr\leq\varepsilon_{\nu,\,i+1}$.
Given this assumption and both $\Phi_0(\varepsilon_\nu,\varepsilon_\nu\pr)$ and
$\Phi_1(\varepsilon_\nu,\varepsilon_\nu\pr)$ at a given $T$ and $\eta_{n,p}$,
we tabulate the following integrals:
$$\int_{\varepsilon_{\nu,\,i}}^{\varepsilon_{\nu,\,i+1}} d\varepsilon_\nu\pr \varepsilon_\nu^{\prime\,2}
\Phi_{l}(\varepsilon_\nu,\varepsilon_\nu\pr),$$
$$\int_{\varepsilon_{\nu,\,i}}^{\varepsilon_{\nu,\,i+1}} d\varepsilon_\nu\pr \varepsilon_\nu^{\prime\,2}
\varepsilon_\nu\pr\Phi_{l}(\varepsilon_\nu,\varepsilon_\nu\pr),$$
$$\int_{\varepsilon_{\nu,\,i}}^{\varepsilon_{\nu,\,i+1}} d\varepsilon_\nu\pr \varepsilon_\nu^{\prime\,2}
e^{-\beta\omega}\Phi_{l}(\varepsilon_\nu,\varepsilon_\nu\pr),$$
and
$$\int_{\varepsilon_{\nu,\,i}}^{\varepsilon_{\nu,\,i+1}} d\varepsilon_\nu\pr \varepsilon_\nu^{\prime\,2}
\varepsilon_\nu\pr\,e^{-\beta\omega}\Phi_{l}(\varepsilon_\nu,\varepsilon_\nu\pr),$$
where $l=0,1$ and each $i$ is an individual energy group, set up at the beginning
of the calculation.  The integrals over $\varepsilon_\nu\pr$ are computed using 16-point Gauss-Legendre
quadrature, each nested with another 16-point Gauss-Legendre quadrature for the integral over
$\cos\theta$ necessary for each $\Phi_l(\varepsilon_\nu,\varepsilon_\nu\pr)$. These integrals
are tabulated at 30 temperature and 30 $\eta_{n,p}$ points, analogous to the case for neutrino-electron
scattering.  For example, the integral
\begin{eqnarray}
\int_0^\infty d\varepsilon_\nu\pr \varepsilon_\nu^{\prime\,2} \tilde{J}_\nu\pr
\Phi_0(\varepsilon_\nu,\varepsilon_\nu\pr)&=&\sum_{i=1}^{n_f}
A_i\int_{\varepsilon_{\nu,\,i}}^{\varepsilon_{\nu,\,i+1}} d\varepsilon_\nu\pr \varepsilon_\nu^{\prime\,2}
\varepsilon_\nu\pr\Phi_{0}(\varepsilon_\nu,\varepsilon_\nu\pr) \nonumber \\
&+&\sum_{i=1}^{n_f}B_i\int_{\varepsilon_{\nu,\,i}}^{\varepsilon_{\nu,\,i+1}}
d\varepsilon_\nu\pr \varepsilon_\nu^{\prime\,2}
\Phi_{0}(\varepsilon_\nu,\varepsilon_\nu\pr),
\label{expnucsum}
\end{eqnarray}
where $A_i=(\tilde{J}_\nu^{i\,\prime}-\tilde{J}_\nu^{i+1\,\prime})/
(\varepsilon_\nu^{i\,\prime}-\varepsilon_\nu^{i+1\,\prime})$ and
$B_i=\tilde{J}_\nu^{i\,\prime}-A_i\varepsilon_\nu^{i\,\prime}$.
The total source and sink are given by integrals analogous to eq.~(\ref{expnucsum})
with appropriate changes to $A_i$ and $B_i$, depending on if the term in the source or
sink is over $\tilde{J}_\nu\pr$ or $\tilde{H}_\nu\pr$.
In practice, for a given temperature and density, we calculate the necessary integrals
at the four nearest neighbor $T-\eta_{n,p}$ points saved in the table and then interpolate the solution
using a four-point bivariant interpolation scheme.
In this scheme, the energy integral over the kernel is
reproduced extremely well and the primary uncertainty in calculating the total source and sink
is due to the linear interpolation of $\tilde{J}_\nu\pr$ and $\tilde{H}_\nu\pr$ -- the
same as in the neutrino-electron scattering case.  Also interesting is that because
the neutrino-nucleon scattering kernel is so sharply peaked around $\varepsilon_\nu$
and drops off so quickly with $\varepsilon_\nu\pr$, most terms in the sum in eq.~(\ref{expnucsum})
are zero.  When we fill the table, we note the index $i$ of the lowest and highest
$\varepsilon_{\nu,i}\pr$ intervals that contribute significantly (to a part in $10^4$) to the total
$\varepsilon_{\nu}\pr$  integral over $\Phi_l$.  Typically, only four to five energy groups
must be included in the final sum.  This decreases both the size of the table and the
amount of time needed to calculate $\tilde{\eta}_\nu^{\rm scatt}$ and
$\tilde{\chi}_\nu^{\rm scatt}$.  Although the vector and axial-vector couplings
for neutrino-neutron scattering are independent of neutrino species, we calculate separate
tables for $\nu_e$, $\bar{\nu}_e$, and $\nu_\mu$ so as to allow for arbitrary energy grouping
for each species.

{\bf The Elastic Limit:} With the inelastic formalism in hand, it is instructive
to construct the elastic limit. Note that in the limit of zero energy transfer,
$S(q,\omega)\rightarrow S(0)=2\pi n_{n,p}\,\delta(\omega)$, where 
$n_{n,p}$ is the neutron or proton number density (see eq. \ref{b6}).
The scattering term can then be written as
\begin{eqnarray}
{\cal L}_{(0)}^{\rm \,scatt}[f_\nu]&=&(1-f_\nu)\frac{Cn_{n,p}}{c(2\pi\hbar c)^3}
\int_0^\infty d\varepsilon_\nu^\prime\varepsilon_\nu^{\prime\,2}
e^{-\beta\omega}\delta(\omega)\int_{-1}^{+1}d\mu^{\prime}f_\nu^\prime\int_0^{2\pi}d\phi^{\prime}\,M
\hspace*{1cm}\nonumber \\
&-&f_\nu\frac{Cn_{n,p}}{c(2\pi\hbar c)^3}\int_0^\infty d\varepsilon_\nu^\prime\varepsilon_\nu^{\prime\,2}
\delta(\omega)\int_{-1}^{+1}d\mu^{\prime}(1-f_\nu^\prime)\int_0^{2\pi}d\phi^{\prime}\,M,
\end{eqnarray}
where $M=[(1+\cos\theta)V^2+(3-\cos\theta)A^2]$.
Using eqs.~(\ref{costheta}), (\ref{jtilde}), and (\ref{htilde}) we find that
\begin{equation}
{\cal L}_{(0)}^{\rm \,scatt}[f_\nu]=\frac{4\pi Cn_{n,p}}{c(2\pi\hbar c)^3}
\int_0^\infty d\varepsilon_\nu^\prime
\varepsilon_\nu^{\prime\,2}\delta(\omega)[(1-f_\nu)e^{-\beta\omega}\Xi_{\rm in}
-f_\nu\Xi_{\rm out}],
\end{equation}
where $\Xi_{\rm in}=V^2(\tilde{J}_\nu^\prime+\mu\tilde{H}_\nu^\prime)+
3A^2(\tilde{J}_\nu^\prime-\mu\tilde{H}_\nu^\prime/3)$ and
$\Xi_{\rm out}=V^2(1-\tilde{J}_\nu^\prime-\mu\tilde{H}_\nu^\prime)+
3A^2(1-\tilde{J}_\nu^\prime+\mu\tilde{H}_\nu^\prime/3)$.
Integrating over $\varepsilon_\nu^\prime$ using the delta function, we have that
\begin{eqnarray}
{\cal L}^{\rm \,scatt}_{(0)}[f_\nu]&\,\,=\,\,&4\pi C\varepsilon_\nu^2(V^2+3A^2)
\left[(\tilde{J}_\nu-f_\nu)+\left(\frac{V^2-A^2}{V^2+3A^2}\right)
\mu\tilde{H}_\nu\right] \nonumber \\
&\,\,=\,\,&\sigma_{n,p} n_{n,p}\left[(\tilde{J}_\nu-f_\nu)+\delta_{n,p}\mu\tilde{H}_\nu\right]
\label{pure}
\end{eqnarray}
where $\sigma_{n,p}=(G^2/\pi c)\varepsilon_\nu^2(V^2+3A^2)$, $\delta_{n,p}$ is
the scattering asymmetry for neutrino-neutron or neutrino-proton
scattering,  $G$ is the weak coupling constant, and $c$ is the
speed of light.  The result presented in eq.~(\ref{pure}) is to be compared
with the full scattering part of the collision term:
\begin{equation}
-\kappa_sf_\nu+
\frac{\kappa_s}{4\pi}\int\Phi({\bf \Omega},{\bf \Omega\pr})\,f_\nu({\bf \Omega\pr})\,d\Omega\pr.
\end{equation}
Performing a first-order Legendre series expansion of the integral elastic scattering source term,
and combining the scattering sink, $-\kappa_sf_\nu$,
with the scattering source, $\kappa_s \tilde{J}_\nu+{\kappa_s\delta}\mu\tilde{H}_\nu$,
we obtain eq.~(\ref{pure}) and an expression for $\delta_{n,p}$.

Note that taking the zeroth moment of ${\cal L}_{(0)}^{\rm \,scatt}[f_\nu]$
yields zero, as it should in the elastic scattering limit.
Further, note that the first moment is non-zero:
\begin{eqnarray}
\frac{1}{2}\int_{-1}^{+1} d\mu \mu\,{\cal L}_{(0)}^{\rm \,scatt}[f_\nu]&=&
\frac{G^2}{\pi c}n_{n,p}\varepsilon_\nu^2\left[\frac{1}{3}\tilde{H}_\nu
(V^2-A^2)-\tilde{H}_\nu(V^2+3A^2)\right] \nonumber \\
&=&-\frac{G^2}{\pi c}n_{n,p}\varepsilon_\nu^2\tilde{H}_\nu(V^2+3A^2)
\left[1-\frac{1}{3}\delta_{n,p}\right] \nonumber \\
&=&-\sigma_{n,p} n_{n,p}\tilde{H}_\nu
\left[1-\frac{1}{3}\delta_{n,p}\right] \, .
\end{eqnarray}
This expression is to be compared with the elastic scattering momentum
source term on the right-hand side of the first moment of the transport
equation.  In fact, the quantity $\sigma_{n,p} n_{n,p}(1-\delta_{n,p}/3)$
defines the elastic transport cross section (see eq. \ref{transcs}).

\section{Dynamic Structure Factors for Neutrino--Nucleon Interactions}
\label{dynamsf}

An alternate formalism for handling inelastic neutrino-nucleon scattering that
more straightforwardly than in \S\ref{nunucleon} generalizes to include nucleon-nucleon correlations,
whether due to fermi blocking or nuclear interactions, involves the dynamical structure factor $S(q,\omega)$.
Our discussion here follows closely that found in Burrows and Sawyer (1998,1999).
Recent explorations into the effects of many--body correlations on neutrino--matter opacities
at high densities have revealed that for densities above $10^{14}$ gm cm$^{-3}$
both the charged--current and the neutral--current interaction rates are decreased
by a factor of perhaps 2 to 3, depending on the density and the 
equation of state (Burrows \& Sawyer 1998,1999; Reddy, Prakash, \& Lattimer 1998; Yamada 1999).
Furthermore, it has been shown that
the rate of energy transfer due to neutral--current scattering off of nucleons
exceeds that due to $\nu_\mu$--electron scattering (Janka et al. 1996; 
Thompson, Burrows, \& Horvath 2000).  Previously, it had been assumed that
neutrino--nucleon scattering was elastic (Lamb \& Pethick 1977).  However,
these recent reappraisals reveal that the product of the underestimated
energy transfer per neutrino--nucleon scattering with cross section exceeds the corresponding
quantity for neutrino--electron scattering.  Since $\nu_e$ and $\bar{\nu}_e$ neutrinos  participate
in super--allowed charged--current absorptions on nucleons,
neutrino--nucleon scattering has little effect on their rate of equilibration.  However, such
scattering is important for $\nu_\mu$ and $\nu_\tau$ equilibration.
Since the many--body correlation suppressions appear only above neutrinosphere
densities ($\sim 10^{11} - 10^{13}$ gm cm$^{-3}$), it is only
the kinematic effect, and not the interaction effect, that need be considered
when studying the emergent spectra.
In the following we adhere closely to the approach and formal development in Burrows and Sawyer (1998,1999).
Without interactions, the relevant dynamical structure factor, $S(q,\omega)$,
for neutrino--nucleon scattering is simply

\begin{equation}
S(q,\omega)= 2\int \frac{d^3p}{(2\pi)^3}\f(|{\bf  p}| )
(1-\f(|{\bf p+q}|))2\pi\delta(\omega+\epsilon_{\bf p}-\epsilon_{\bf p+q})\, ,
\label{b4}
\end{equation}
where $\f({\bf |p|})$ is the nucleon Fermi--Dirac distribution function,
$\epsilon_{\bf p}$ is the nucleon energy, $\omega$ is the energy transfer to the medium,
and ${\bf q}$ is the momentum transfer.
The magnitude of ${\bf q}$ is related to $\omega$ and $E_1$,
the incident neutrino energy, through the neutrino
scattering angle, $\theta$, by the expression (see text in \S\ref{section:inelastic} after eq. \ref{costheta}),
\begin{equation}
q=[E_1^2+(E_1-\omega)^2-2E_1(E_1-\omega) \cos\theta]^{1/2} \, .
\end{equation}
In the elastic limit and ignoring final--state nucleon blocking,
$S(q,\omega)=2\pi\delta(\omega)n_n$, the expected result,
where $n_n$ is the nucleon's number density.

The neutral current scattering rate off of either neutrons or protons is (Burrows \& Sawyer 1998),
\beq
\frac{d^2\Gamma}{d\omega d\cos\theta}=
(4\pi^2)^{-1}G^2_W(E_1-\omega)^2
[1-\f_\nu(E_1-\omega)]{\cal{I_{\rm NC}}} \, ,
\eeq
where
\beq
{\cal{I_{\rm NC}}}=
\Bigl[(1+\cos\theta)V+(3-\cos\theta)A\Bigr]S(q,\omega)
\eeq
and
\beqa
S(q,\omega)&=&2\rm{Im}\Pi^{(0)}(1-e^{-\beta\omega})^{-1}\, .
\label{structf}
\eeqa
$V$ and $A$ are the
applicable vector and axial--vector coupling terms (see \S\ref{nupro} and \S\ref{nuneu}) and $\beta = 1/kT$.
The free polarization function, $\rm \Pi^{(0)}$, contains the full kinematics of the
scattering, as well as blocking due to the final--state nucleon, and
the relevant imaginary part of $\rm \Pi^{(0)}$ is given by:
\beq
{\rm{Im}}\Pi^{(0)}(q,\omega)=\frac{m^2}{2\pi q\beta}
\log\left[\frac{1+e^{-Q_+^2+\beta\mu}}{1+e^{-Q_+^2+\beta\mu-\beta\omega}}\right]\, ,
\label{structim}
\eeq
where
\beq
Q_\pm=\left(\frac{m\beta}{2}\right)^{1/2}\left(\mp\frac{\omega}{q}+
\frac{q}{2m}\right),
\eeq
$\mu$ is the nucleon chemical potential, and $m$ is the nucleon mass.
The dynamical structure factor, $S(q,\omega)$, contains all of the information necessary to
handle angular and energy redistribution due to scattering.  The
corresponding term on the right--hand--side of the
transport equation is:
\beq
{\cal S}[{\cal F}_\nu]=(2\pi)^{-3}G^2_W
\int d^3p_\nu\pr\,
{\cal I}_{{\rm NC}}\, \Xi_{SF}
\nonumber 
\eeq
where 
\beq
\Xi_{SF} = [1-{\cal F}_\nu(E_1)]{\cal F}_\nu^{\prime}(E_1-\omega)
e^{-\beta\omega}-{\cal F}_\nu(E_1)[1-{\cal F}_\nu^{\prime}(E_1-\omega)]
\label{redistribution}
\eeq
and $p_\nu\pr$ is the final state neutrino momentum.

In the non-degenerate nucleon limit, eq. (\ref{structim}) can be expanded to lowest order in $Q_+^2$
to obtain, using eq. (\ref{structf}), an approximation to the dynamical structure factor:

\beqa
S(q,\omega)&=&\frac{n(2\pi m \beta)^{1/2}}{q}\, e^{-Q_+^2}\, ,
\label{structapp}
\eeqa
where $n$ is the nucleon number density.  This says that for a given momentum transfer the dynamical
structure factor is approximately a Gaussian in $\omega$.

For charged--current absorption process, $\enu\,+\,n\, \righta\, e^-\,+\,p$,
${\rm{Im}}\Pi^{(0)}(q,\omega)$ is given by
a similar expression:

\begin{equation}
{\rm Im} \Pi^{(0)}(q,\omega)=\frac{m^2}{2\pi \beta q} \log
\Bigl[\frac{1+e^{-Q_+^2+\beta \mu_n}}{1+e^{-Q_+^2+\beta
\mu_p-\beta\omega}} \Bigr] \, .
\label{cc13}
\end{equation}
Eq. (\ref{cc13}) inserted into eq. (\ref{structf}) with a $(1-e^{-\beta(\omega +\hat{\mu})})$,
as is appropriate for the charged--current process, substituted for $(1-e^{-\beta\omega})$,
results in an expression that is a bit more general than the one employed to date
by most practitioners (Bruenn 1985), {\it i.e.}, $S=(X_n-X_p)/(1-e^{-\hat{\mu}/T})$.
In the non-degenerate nucleon limit, the structure factor for the charged--current process
can be approximated by eq. (\ref{structapp}) with $n=n_n$.
Note that for the structure factor of a charged-current interaction one must
distinguish between the initial-- and the final--state nucleons and, hence,
between their chemical potentials.
To obtain the structure factor for the $\anu_e$ absorption process,
one simply permutes $\mu_n$ and $\mu_p$ in eq. (\ref{cc13}) and substitutes $-\hat{\mu}$ for $\hat{\mu}$
in the $(1-e^{-\beta(\omega +\hat{\mu})})$ term.

In general, the widths of the structure factors 
are larger than might have been expected for 
scattering off of ``heavy'' particles.  This is because in the past
people thought that the neutrino could lose in $\nu$--nucleon scattering
an energy equal to only about $-E_1^{2}/m_nc^2$,
{\it i.e.} that the fractional energy lost is of order $p_\nu/m_nc$ ($\sim$1\%).
However, this assumes that the nucleons are stationary.  In fact, they are
thermal and,  the fractional energy they can in a collision transfer to the
neutrino is of order $p_n/m_nc$.  Since the nucleons have such a large mass, if they
and the neutrino have the same energy, $p_n/m_nc$ is much larger than $p_\nu/m_nc$,
at incident neutrino energies of 10--30 MeV by as much as an order of magnitude.
The formalism above incorporates the kinematics of such a collision,
a realistic Fermi--Dirac energy distribution for the nucleons, and final--state nucleon
blocking.  The upshot is broad distributions. 
Including many--body effects further
flattens and broadens the distribution, while lowering the central values of
${d\sigma}/{d\omega}$, as well as the total integral over $\omega$ (Burrows \& Sawyer 1998,1999).

\subsection{Aside: Static Structure Factors}
\label{static}

In the limit of heavy nucleons,
when we perform the integration in eq. (\ref{b4}) over a range of
$\omega$s and evaluated the inner factors at
$\omega=0$, we can express this limit as,

\begin{equation}
(2\pi)^{-1}S(q,\omega)\rightarrow
(2\pi)^{-1}\delta (\omega )\int d\omega' S(q,\omega')\equiv \delta (\omega )S(q) ,
\label{b6}
\end{equation}
where $S(q)$ is the {\it static} structure factor.  
At the high densities and temperatures achieved in the supernova context,
the $\omega = 0$ (elastic) limit is not particularly accurate 
(Burrows \& Sawyer 1998; Reddy, Prakash, \& Lattimer 1998).

$S(q)$ is merely the Fourier transform of the thermally--averaged
density--density correlation function.  This is the classic result that
scattering off of a medium is in reality scattering off of the {\it fluctuations}
in that medium.  Also of interest is the long--wavelength limit, $q \rightarrow 0$,
justified when the neutrino wavelength is much bigger than the
interparticle separation. Statistical mechanics
provides two useful and equivalent expressions
for the long--wavelength limit, $S(0)$, the first (Landau \& Lifschitz 1969)
in terms of the isothermal compressibility of the medium $K_T$
($=-\frac{\partial \log V}{\partial P}\bigr|_T$),

\begin{equation}
S(0)= \bar{n}^2 \beta^{-1} K_T = \bar{n} \frac{K_T}{K_0},
\label{fluct1}
\end{equation}
where $K_0$ is the ideal gas compressibility and
$\bar{n}$ is the average nucleon density,
and the second in terms of the derivative of the density with respect
to the chemical potential of the nucleons, $\mu$,

\begin{equation}
S(0)=  \beta ^{-1}\frac{\partial \bar{n}}{\partial \mu}.
\label{fluct2}
\end{equation}
In the ideal gas limit of no correlation between particles, eqs.~(\ref{fluct1})
and (\ref{fluct2}) show that $S(q)$ is simply equal
to the number density, $\bar{n}$, as expected from
eq.~(\ref{b4}), without blocking.
Eq.~(\ref{fluct1}) reveals that if $K_T$ is small because the
matter is stiff, in the long--wavelength limit
the neutrino--matter cross sections are {\it suppressed}.
In general, when we acknowledge that the neutrino-matter 
interaction has axial--vector current and nucleon isospin terms, we 
require separate correlation functions for the neutron and the proton,
as well as for spin correlations. These depend upon susceptibilities
that are different from the compressibility, but we find
suppression in these terms as well (Burrows \& Sawyer 1998; Reddy, Prakash, \& Lattimer 1998).

Eq.~(\ref{fluct2}), equivalent to eq.~(\ref{fluct1}) by a thermodynamic identity, is
a powerful result of great generality.
In standard approximation schemes for the many--body problem, the
distribution function for a nucleon species is given by the
Fermi--Dirac distribution in which the chemical potential $\mu$ is
replaced by $\mu-v(\bar{n})$, where $v(\bar{n})$ is the average
energy of interaction of the nucleon with the other nucleons
and is a function of the density. Thus, the density is given implicitly by

\begin{equation}
 \bar{n}=2\int \frac{d^3p}{(2\pi)^{3}}[1+e^{\beta[(p^2/(2m)-\mu+v(\bar{n})]}]^{-1}.
\label{a15}
\end{equation}

The expression (\ref{a15}) holds in the Hartree approximation;
it holds in approaches that introduce mean meson fields instead
of nuclear potentials; it holds in the Landau Fermi liquid theory (FLT),
subject to the proviso that we use only results in which the derivative
of the potential $v$ (with respect to the $ \bar{n}$) enters; and
it holds in approaches using the Skyrme potential.

Differentiating (\ref{a15}), we can solve for
$\frac{\partial \bar{n}}{\partial \mu}$ and $S(0)$,

\begin{equation}
S(0)=\beta^{-1}\frac{\partial  \bar{n}}{\partial \mu}=h(\mu)\bigl[1+h(\mu)
\frac{\partial v}{\partial \bar{n}}\bigr]^{-1},
\label{a16}
\end{equation}
where

\begin{equation}
h(\mu)=2 \int \frac{d^3p}{(2\pi)^3}\frac{e^{\beta[p^2/(2m)-\mu+v]}}
{[1+e^{\beta[p^2/(2m)-\mu+v]}]^2}\\
= 2\int \frac{d^3p}{(2\pi)^{3}} \f(p)(1-\f(p))
\label{a17}
\end{equation}
and $\f(p)$ is the Fermi--Dirac function, but
with the chemical potential displaced by $v$.  If we regard
particle densities as inputs to our calculations, then the displacement
of the chemical potential by the nuclear potential is irrelevant, since
the same difference, $\mu-v$, enters the calculation of the density in
terms of the chemical potential. Thus, the numerator of (\ref{a16})
contains no more than the familiar Pauli blocking effects (for the case $q=0$);
the denominator contains all of the effect of the interactions.

As an example, consider a two--nucleon potential $V(r)$. In the Hartree approximation,
the average potential seen by a single nucleon is given by $v=\bar{n}U$, where
$U= \int d^3x V(x)$, and (\ref{a16}) becomes

\begin{equation}
S(0)=h(\mu)\bigl[1+h(\mu)U\bigr]^{-1},
\label{b20}
\end{equation}
the potential providing an enhancement, if negative, and a suppression, if positive.
The latter is the case for high--density nuclear matter.

\subsection{Procedure for Calculating $\nu$--nucleon Structure Functions 
for Neutral-Current Scattering Including Interactions}
\label{many}

Now, taking only the neutron part of the vector--current coupling, the
differential neutrino-nucleon scattering rate is given by,

\begin{eqnarray}
\frac{d^2 \Gamma}{d \omega \hspace{1 pt} d \cos\theta}=
(4 \pi^2)^{-1}G_W^2 E_2^2[1-f_{\nu}(E_2)]\Bigl[\bigl(1+
\cos\theta\bigr)(C_{V}^n)^2S_{nn}(q,\omega)
\nonumber\\
+\bigl(3-\cos\theta\bigr)g_{A}^2\bigl[ S^A_{pp}(q,\omega)+
S^A_{nn}(q,\omega)-2S^A_{pn}(q,\omega)  \bigr],
\label{a38}
\end{eqnarray}
where $E_2$=$E_1-\omega$.

The structure functions, $S$ (Fermi) and $S^A$ (Gamow-Teller; axial), are elements of separate
$2\times2$ symmetric matrices. For the vector dynamic structure
function, $S$, we have

\begin{displaymath}
S(q,\omega)=
\pmatrix{S_{pp}(q,\omega)&S_{pn}(q,\omega)\cr S_{pn}(q,\omega)&S_{nn}(q,\omega)\cr}.
\end{displaymath}

The structure function matrix is given by,

\begin{equation}
S(q,\omega)=2 {\rm Im} \Bigl[\Pi^{(0)}(q,\omega)
[1-v(q) \Pi^{(0)}(q,\omega)]^{-1}
\Bigr](1-e^{-\beta\omega})^{-1}
\label{ebetao}
\end{equation}
where

\begin{displaymath}
\Pi^{(0)}(q,\omega)=
\pmatrix{\Pi^{(0)}_p(q,\omega) &0\cr 0&\Pi^{(0)}_n (q,\omega)\cr}
\end{displaymath}
and $\Pi^{(0)}_p$ and $\Pi^{(0)}_n$ are given by the polarization function
and evaluated with
the proton and neutron chemical potentials, respectively (eq. \ref{structim}).
The potential matrix is,

\begin{displaymath}
v=\pmatrix{v_1+v_2+4\pi e^2 (q^2+q_{TF}^2)^{-1}&v_1-v_2\cr v_1-v_2&v_1+v_2\cr},
\end{displaymath}
where the $v$'s were defined in terms of Fermi liquid parameters and the term containing
$q_{TF}$ is the Thomas-Fermi screened Coulomb potential ($q_{TF}^2=4 e^2\pi^{1/3}(3\bar{n}_p)^{2/3}$).
Following Burrows and Sawyer (1998,1999) and for simplicity, we use here FLT and Landau parameters, 
in lieu of a more developed nuclear interaction model.

In a real calculation, in all the kinematic expressions the nucleon mass ($m$) is to be
replaced by $m^*$. Unfortunately, the relation of Landau parameters to
experimental results depends upon the effective mass in model--dependent
ways. Taking  $m^*=0.75 m_n$ as our fiducial value for the effective mass,
we use parameters from Backman, Brown, \& Niskanen (1985) and 
Brown and Rho (1981): $F_0= -0.28; F_0'=0.95; G_0=0;
G_0'=1.7$ and $\lambda=2.63 \times10^{-5} {\rm MeV}^{-2}$, obtaining,

\begin{eqnarray}
&v_1=-7.4 \times 10^{-6}\,{\rm MeV}^{-2}
\nonumber\\
&v_2=2.5\times 10^{-5}\,{\rm MeV}^{-2}
\nonumber\\
&v_3=0
\nonumber\\
&v_4=4.5\times 10^{-5}\,{\rm MeV}^{-2}.
\label{asabove}
\end{eqnarray}

For other values of the effective mass, we keep these potentials at the
same value, which is to say we assume that the Landau parameters are
proportional to  $m^*/ m$. 

The form for the Gamow--Teller matrix, $S^A(q,\omega)$, is the same as that for $S$,
except that the potential matrix is replaced by $v^A$

\begin{displaymath}
v^A=\pmatrix{v_3+v_4&v_3-v_4\cr v_3-v_4&v_3+v_4\cr}.
\end{displaymath}

Taking the matrix inverses leads to the following forms for the
combinations of structure functions that appear in (\ref{a38})

\begin{equation}
S_{nn}(q,\omega)=2 {\rm Im} \bigl[\Pi_n^{(0)}D_V^{-1}\bigl](1-e^{-\beta\omega})^{-1},
\label{vector}
\end{equation}
where

\begin{equation}
D_V=1-(v_1+v_ 2)\Pi_n^{(0)}-(v_1-v_2)^2\Pi_n^{(0)}\Pi_p^{(0)}{Q_V}^{-1}\, .
\label{DVterm}
\end{equation}
$Q_V$ is given by the expression:  
\begin{equation}
Q_V=1-4\pi e^2(q^2+q_{TF}^2)^{-1}\Pi_p^{(0)}-(v_1+v_2)\Pi_p^{(0)}\, .
\label{DVterm2}
\end{equation}

If, as in (\ref{asabove}), we take $v_3 =0$, we obtain
the simple result for the axial--current terms,

\begin{equation}
S_A(q,\omega)=
2 {\rm Im} \Big[\frac{\Pi_p^{(0)}(q,\omega)+\Pi_n^{(0)}(q,\omega)}
{1-v_4[\Pi_p^{(0)}(q,\omega)+\Pi_n^{(0)}(q,\omega)]}\Bigr](1-e^{-\beta\omega})^{-1}\, .
\label{axial}
\end{equation}

For the Fermi term, since $ C_V^{(p)}=1/2-2\sin^2\theta_W\sim 0 $,
we drop the proton structure function in
(\ref{a38}).  Furthermore, we use the potential parameters
given in eq.~(\ref{asabove}), and in eq.~(\ref{DVterm})
we drop the third term.  This term would have
been significant had it not been for the
Coulomb term in the denominator, an illustration of the importance
of the explicit inclusion of Coulomb forces, even for the neutron
density correlations.
Since the $v_i$s are all real,
we obtain for the structure factors used in (\ref{a38}),

\begin{equation}
S_{F}(q,\omega)=2 {\rm Im} \Pi_n^{(0)}(1-e^{-\beta\omega})^{-1}{\cal C_V}^{-1},
\label{vectel}
\end{equation}
where
\begin{equation}
{\cal C_V}= (1 - v_F{\rm Re}\Pi_n^{(0)})^2 +
v_F^{2}({\rm Im}\Pi_n^{(0)})^2,
\label{vectel2}
\end{equation}
and

\begin{equation}
S_A(q,\omega)=
2 \Bigl[{\rm Im}\Pi_p^{(0)}(q,\omega)+{\rm Im}\Pi_n^{(0)}(q,\omega)\Bigr]
(1-e^{-\beta\omega})^{-1}{\cal C_A}^{-1},
\label{axel}
\end{equation}
where
\begin{equation}
{\cal C_A}={\cal C}_{{\cal A}1} + {\cal C}_{{\cal A}2}\, .
\label{axel2}
\end{equation}
${\cal C}_{{\cal A}1}$ and ${\cal C}_{{\cal A}2}$ are given by the expressions:
\begin{equation}
{\cal C}_{{\cal A}1}=
\Bigl[1-v_{GT}({\rm Re}\Pi_p^{(0)}(q,\omega)+{\rm Re}\Pi_n^{(0)}(q,\omega))\Bigr]^2
\label{axel3}
\end{equation}
and 
\begin{equation}
{\cal C}_{{\cal A}2}=
v_{GT}^{2}\Bigl[{\rm Im}\Pi_p^{(0)}(q,\omega)+{\rm Im}\Pi_n^{(0)}(q,\omega)\Bigr]^2 \, .
\label{axel4}
\end{equation}

The $F$ in $ S_{F}(q,\omega) $ and the $A$ in $ S_A(q,\omega) $ stand for Fermi
and Gamow--Teller (axial) and $v_F$ and $v_{GT}$ equal
$(v_1+v_ 2)$ and $v_4$, respectively, in Fermi Liquid Theory.  $S_A(q,\omega)$ in eq. (\ref{axel} is
now the entire axial term in eq. (\ref{a38}).  ${\cal C_{V,A}}$ is the correction factor due to
many--body effects for a given momentum transfer (or scattering angle) and energy
transfer.  A similar procedure is employed for calculating the many-body corrections
to the charged-current rates (Burrows \& Sawyer 1999).

\subsection{Collective Excitations of the Medium}

\begin{figure}
\vspace*{6.0in}
\hbox to\hsize{\hfill\includegraphics{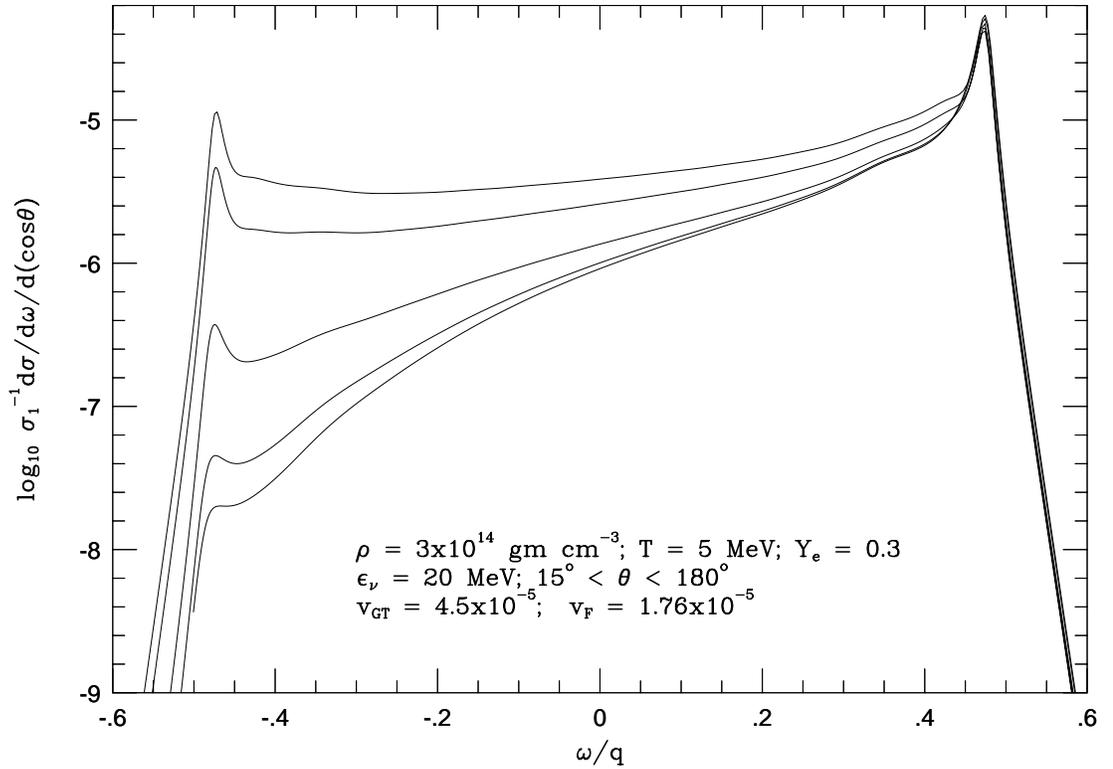}\kern+6in\hfill}
\caption{Log$_{10}$ of the doubly--differential cross section
for neutral--current neutrino--nucleon
scattering versus $\omega/q$ for scattering angles 15$^\circ$,
45$^\circ $, 90$^\circ $, 135$^\circ $, and 180$^\circ $ .
The calculations were performed at a temperature of 5 MeV, a $Y_e$ of 0.3, a $\rho$
of $3\times10^{14}$ g cm$^{-3}$, and an incident neutrino energy of 20 MeV.
The default potentials  ($v_{GT}=4.5\times10^{-5}$
and $v_F=1.76\times10^{-5}$) 
and effective mass ($m^*=0.75\, m_n$) were employed.
The differential cross section is divided by the total
scattering cross section ($\sigma_1$) in the non--interacting,
no--nucleon--blocking,
$\omega=0$ limit.
(Figure taken from Burrows and Sawyer 1998.)
}
\label{figBS1}
\end{figure}

\begin{figure}
\vspace*{6.0in}
\hbox to\hsize{\hfill\includegraphics{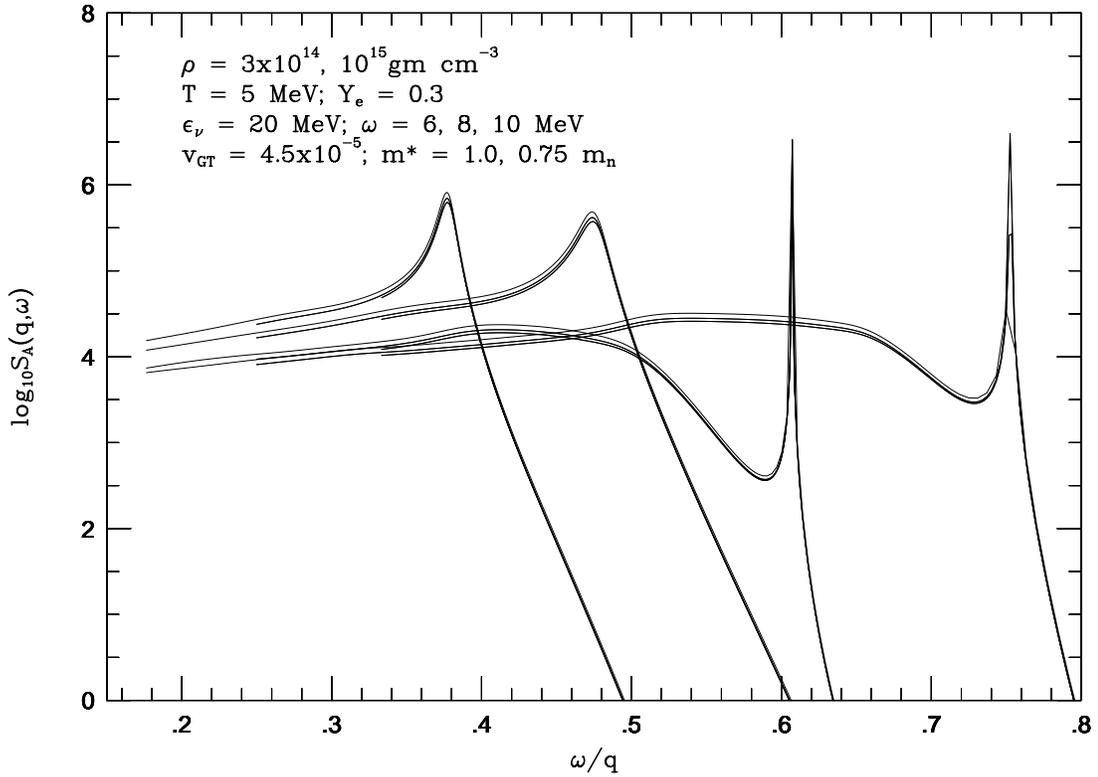}\kern+6in\hfill}
\caption{Log$_{10}$ of the Gamow--Teller structure function
versus $\omega/q$ for an incident neutrino energy
of 20 MeV, energy transfers, $\omega$, of 6, 8,
and 10 MeV, two values of the effective
mass ($ m^* = [0.75m_n, 1.0 m_n]$) and two values
of the density ($\rho = 3\times10^{14}$ and $10^{15}$ g cm$^{-3}$).
A temperature of 5 MeV and a $Y_e$ of 0.3 were used,
as was the default $v_{GT}$ ($=4.5\times10^{-5}$). 
(Figure taken from Burrows and Sawyer 1998.)
}
\label{figBS2}
\end{figure}

Following Burrows and Sawyer (1998,1999), we note 
that for most regions of phase space, ${\cal C_V}$ and 
${\cal C_A}$ in eqs. (\ref{vectel2}) and (\ref{axel2}) are greater
than one and represent suppression in the
scattering rates.  Their effects on the integrals
over $\omega$ and $\theta$ are always
suppressive.  However, the terms containing
the real parts have roots; these roots represent
collective excitations.  For the Fermi term, zero
sound in the medium can be generated if the
scattering has a ($\omega,q$) pair that satisfies
the mode's dispersion relation, {\it i.e.}, if
it hits the resonance. Similarly, for the Gamow--Teller
term, spin waves in the protons and the neutrons
(related by a set phase) can be generated.   These
modes are the traveling--mode equivalents of the
Gamow--Teller resonance in nuclei (a standing wave).
The zero sound of the Fermi part
is analogous to the Giant--Dipole resonance in nuclei.
The resonances increase the structure function when
the scattering transfer ratio, $\omega/q$, equals
the ratio of the collective excitation's angular frequency
and wavenumber.  For a given scattering angle, one
can plot the differential cross section in $\omega$
and $\cos\theta$ as a function of $\omega/q$ to see
the resonances.  In Figure~(\ref{figBS1}), we display this
for five different angles between 15$^\circ $ and 180$^\circ $, an
incident neutrino energy of 20 MeV, a temperature
of 5 MeV, a density of $3\times10^{14}$ g cm$^{-3}$,
and an electron fraction, $Y_e$, of 0.3.
We see in Figure~\ref{figBS1} that the resonances in
both the forward and the backward directions line
up at the same values of $\omega/q$, as expected
for a collective mode, and we can straightforwardly
calculate the mode's dispersion relation.
This is akin to the \v{C}erenkov effect.
Note that the Gamow--Teller term dominates the Fermi term,
so that in Figure~\ref{figBS1} we are really seeing the
spin waves related to the Gamow--Teller resonance.  However, the
dispersion relations for zero sound and these
spin waves are generally similar.  In fact,
recalling the classic result of Fetter and Walecka (1971) that in the weak--coupling limit,
the speed of zero sound in a degenerate system is $\sim v_{fermi}$, where
$v_{fermi}$ is the Fermi velocity, and recalling that for nucleons
in nuclei $v_{fermi}$ is $\sim0.3c$, the calculated
resonance value of $\omega/q$ is not unexpected.
In Figure~\ref{figBS2}, we plot the Gamow--Teller
structure function versus $\omega/q$ for various
values of $\omega$, $m^*$, and two values of the density.  At $m^*=m_n$,
for each value of the density we obtain a sharp
resonance, but at two different speeds,
reflecting the crude $\rho^{1/3}$--dependence
expected for $v_{fermi}$.  For a given density,
the mode speed is seen in Figure~\ref{figBS2} to be
inversely proportional to the effective mass.
The width of the resonance is determined by the magnitude
of the imaginary part of the polarization function.

\section{$e^+e^-$ Annihilation}
\label{eplus}    

Ignoring phase space blocking of neutrinos in the final state and taking
the relativistic limit ($m_e\rightarrow 0$), the total electron--positron
annihilation rate into neutrino--antineutrino pairs can
be written in terms of the electron and positron phase space densities (Dicus 1972):
\beq
Q_{\nu_e\bar{\nu}_e}=
K_i\left(\frac{1}{m_ec^2}\right)^2\left(\frac{1}{\hbar c}\right)^6
\int\int
\f_{e^-}\f_{e^+}(\varepsilon_{e^-}^4\varepsilon_{e^+}^3+\varepsilon_{e^-}^3
\varepsilon_{e^+}^4)\,d\varepsilon_{e^-}\,d\varepsilon_{e^+}\,\,,
\label{dicusrate2}
\eeq
where $K_i=(1/18\pi^4)c\sig_o(C_V^2+C_A^2)$.  Again, $C_V=1/2+2\sinw$ for
electron types, $C_V=-1/2+2\sinw$ for mu and tau types, and
$C_A^2=(1/2)^2$. Rewriting eq. (\ref{dicusrate2}) in terms of the Fermi
integral $F_n(\eta)$, we obtain:
\beq
Q_{\nu_e\bar{\nu}_e}=K_i\,(kT)\left(\frac{kT}{m_ec^2}\right)^2
\left(\frac{kT}{\hbar c}\right)^6
\left[F_4(\eta_e)F_3(-\eta_e)+F_4(-\eta_e)F_3(\eta_e)\right] \,\, ,
\eeq
where $\eta_e\equiv\mu_e/kT$ and
\beq
F_n(\eta)\equiv\int_0^\infty\,\frac{x^n}{e^{x-\eta}+1}\,dx\,\,.
\eeq
Integrating eq. (\ref{dicusrate2}), we obtain
\beq
Q_{\nu_e\bar{\nu}_e}\simeq9.7615\times 10^{24}\,\left[\frac{kT}{{\rm
MeV}}\right]^9\,f(\eta_e)\,\, {\rm ergs\, cm^{-3} s^{-1}} \, ,
\label{pairtotal}
\eeq
where
\beq
f(\eta_e)=
\frac{F_4(\eta_e)F_3(-\eta_e)+F_4(-\eta_e)F_3(\eta_e)}{2F_4(0)F_3(0)}\,\, .
\eeq
For $\nu_\mu\bar{\nu}_\mu$ and $\nu_\tau\bar{\nu}_\tau$ production
combined,
\beq
Q_{\nu_{\mu,\tau}\bar{\nu}_{\mu,\tau}}\simeq 4.1724\times
10^{24}\,\left[\frac{kT}{{\rm MeV}}\right]^9\,f(\eta_e)\,\,
{\rm ergs\, cm^{-3} s^{-1}} \, .
\label{enumurate}
\eeq

One can easily derive the spectrum of the total radiated neutrino energy ($\varepsilon_{T}$)
by inserting a delta function ($\int \delta (\varepsilon_{T}-\varepsilon_{e^-}-\varepsilon_{e^+}) d\varepsilon_{T}$)
into eq. (\ref{dicusrate2}).   Recall that the total energy of the neutrinos in the final
state is equal to the sum of the electron and positron energies in the initial state.
Integrating first over $\varepsilon_{e^+}$ to annihilate
the delta function and then over $\varepsilon_{e^-}$ to leave a function of $\varepsilon_{T}$,
one obtains:

\beq
\frac{dQ}{d\varepsilon_{T}}=
K_i\left(\frac{1}{m_ec^2}\right)^2\left(\frac{1}{\hbar c}\right)^6
\int_{0}^{\varepsilon_{T}} \varepsilon_{T} (\varepsilon_{T}-\varepsilon_{e^-})^3 \varepsilon_{e^-}^3
\f_{e^-}[\varepsilon_{e^{-}}] \f_{e^+}[\varepsilon_{T}-\varepsilon_{e^{-}}] \,d\varepsilon_{e^-}\,\, .
\label{dicusrate_et}
\eeq
The numerical evalution of eq. (\ref{dicusrate_et}) is straightforward.  The average of $\varepsilon_{T}$
is equal to:

\beq
\langle \varepsilon_{T} \rangle = \Bigl(\frac{F_4(\eta_e)}{F_3(\eta_e)} + \frac{F_4(-\eta_e)}{F_3(-\eta_e)}\Bigr)T\, ,
\label{averageet}
\eeq
which near $\eta_e \sim 0$ is $\sim 8T$ and for $\eta_e >> 1$ is $\sim 4T(1+\eta_e/5)$.

However, while the total energy loss rate (eq. \ref{pairtotal}) and the spectrum of $\varepsilon_{T}$
pose no great mathematical problems, the production spectrum of an individual neutrino is not so easily reduced to a
simple integral or to an analytic expression.  This is due primarily to the awkward
integration of the angular phase space terms, while subject to the momentum conservation
delta function, and to the explicit dependence of the matrix elements
on the electron/neutrino angles.
>From \cite{dicus72}, averaging over initial states and summing over final states, the
matrix element for the $e^+e^-\to\nu\bar{\nu}$ process in the $m_e=0$ limit is:
\beq
\frac{1}{4}\sum_s|{\cal {M}}|^2=16G^2[(C_V+C_A)^2{\bf p \cdot
q}_{\bar{\nu}}\,{\bf p}\pr\cdot{\bf q}_\nu+(C_V-C_A)^2{\bf p\cdot
q}_\nu\,{\bf p}\pr\cdot{\bf q}_{\bar{\nu}}]\,\, ,
\label{avematrix}
\eeq
where $p$ and $p\pr$ are the four-momenta of the electron and positron,
respectively, and $q_\nu$ and $q_{\bar{\nu}}$ are the four-momenta of the
neutrino and antineutrino, respectively.  Using the formalism of
\cite{bruenn_1985} and
Fermi's Golden rule, expanding the production kernel in the traditional truncated
Legendre series, performing the trivial angular integrals, taking the
non--trivial angular integrals from \cite{bruenn_1985}, and ignoring final--state
neutrino blocking, we obtain for the single--neutrino
source spectrum due to $e^+e^-$ annihilation:
\beq
\frac{dQ}{d\varepsilon_\nu}=\,\frac{8\pi^2}{(2\pi \hbar
c)^6}\,\varepsilon_\nu^3\,\int_0^\infty\,d\varepsilon_{\bar{\nu}}\,
\varepsilon_{\bar{\nu}}^2\,\Phi_0^p(\varepsilon_\nu,\varepsilon_{\bar{\nu}})\,\, ,
\label{qspectrum}
\eeq
where
\beq
\Phi_0^p(\varepsilon_\nu,\varepsilon_{\bar{\nu}})=\frac{G^2}{\pi}
\int_0^{\varepsilon_\nu+\varepsilon_{\bar{\nu}}}d\varepsilon_{e^-}{\cal
{F}}_{e^-}[\varepsilon_{e^-}]{\cal
{F}}_{e^+}[\varepsilon_\nu+\varepsilon_{\bar{\nu}}-\varepsilon_{e^-}]\,
H_0(\varepsilon_\nu,\varepsilon_{\bar{\nu},}\varepsilon_{e^-})\,\,,
\label{phi0}
\eeq
and
\beq
H_0(\varepsilon_\nu,\varepsilon_{\bar{\nu}},\varepsilon_{e^-})=(C_V+C_A)^2\,
J_0^I(\varepsilon_\nu,\varepsilon_{\bar{\nu}},\varepsilon_{e^-})+(C_V-C_A)^2
\,J_0^{II}(\varepsilon_\nu,\varepsilon_{\bar{\nu}},\varepsilon_{e^-})\,\, .
\label{polynomials}
\eeq
The $J_0$s in eq. (\ref{polynomials}) come from the more obdurate angular
integrals required by the dot products in eq. (\ref{avematrix}) and the momentum
delta function and have the symmetry:
\beq
J_0^I(\varepsilon_\nu,\varepsilon_{\bar{\nu}},\varepsilon_{e^-})=J_0^{II}
(\varepsilon_{\bar{\nu}},\varepsilon_\nu,\varepsilon_{e^-})\,\, .
\eeq
>From eqs. (\ref{qspectrum}) and (\ref{polynomials}), we see that the differences
between the spectra of the $\nu_e$ and $\nu_{\mu}$ neutrinos
flow solely from their correspondingly different values of $(C_V+C_A)^2$ and $(C_V-C_A)^2$.
One can use 4--point Gauss--Legendre integration to calculate eq. (\ref{phi0}) and 16--point
Gauss--Laguerre integration to calculate eq. (\ref{qspectrum}).

At small $\eta_e$, the $e^+ e^-$ annihilation spectra and total energy loss rates for
the $\nu_e$ and $\bar{\nu}_e$ neutrinos are similar, as are the
average emitted  $\nu_e$ and $\bar{\nu}_e$ neutrino energies.  However,
as $\eta_e$ increases, both the total energy radiated in $\bar{\nu}_e$
neutrinos and the average $\bar{\nu}_e$ energy start to lag the corresponding quantities for the $\nu_e$ neutrinos.
This is true despite the fact that the total number of $\nu_e$ and $\bar{\nu}_e$ neutrinos radiated
is the same. If final--state blocking is
ignored, $\langle \varepsilon_i \rangle/T$ is a function of $\eta_e$ alone, becoming linear with $\eta_e$
at high $\eta_e$ and one half of eq. (\ref{averageet}) ($\sim$4.0) at low $\eta_e$.
Note also that $\langle \varepsilon_{\nu_{\mu}} \rangle/T$ and $\langle \varepsilon_{\bar{\nu}_{\mu}} \rangle/T$
are closer to one another than are $\langle \varepsilon_{\nu_e} \rangle/T$ and $\langle \varepsilon_{\bar{\nu}_e} \rangle/T$.
The individual production spectra vary in peak strength, in peak energy, and
in low--energy shape, but they are quite similar on the high--energy tail.  Due to the parity--violating
matrix element for the $e^+ e^- \rightarrow \nu \bar{\nu}$ process and the fact that $\eta_e$ is positive,
the antineutrino spectra of all species are softer than the neutrino spectra.
The pair sums of the integrals under these curves are given by eqs. (\ref{pairtotal}) and (\ref{enumurate}).
For $\eta_e = 0$, 50\% of the pair energy emission of electron types is in $\bar{\nu}_e$ neutrinos, but
at $\eta_e = 10$ only 42\% of this total energy is in $\bar{\nu}_e$ neutrinos.  However, at $\eta_e = 10$,
the $\bar{\nu}_{\mu}$ neutrinos still constitute 48.5\% of the $\nu_{\mu} / \bar{\nu}_{\mu}$ pair emission.
These differences reflect differences in the corresponding coupling constants $C_V$ and $C_A$.

\section{$\nu_i\anu_i$ Annihilation}
\label{paira}  

In the limit of high temperatures and ignoring electron phase space
blocking, the  $\nu_i\anu_i$ annihilation rate into $e^+e^-$ pairs can be
written (Janka 1991):
\beq
Q_{\nu_i\anu_i}=
4K_i\pi^4\left(\frac{1}{m_ec^2}\right)\,\left(\frac{4\pi}{c}\right)^2
\int\int\,\Phi^{\prime}\,J_{\nu_i}J_{\bar{\nu}_i}(\vep_{\nu_i}+\vep_{\bar{\nu}_i})
\,d\vep_{\nu_i}\,d\vep_{\bar{\nu}_i}\,\, ,
\label{jankarate1}
\eeq
where $J_\nu$ is the zeroth moment of the radiation field,
$\epnu$ is the neutrino energy, $K_i$ is defined as before
({\it i.e.,} $K_i=(1/18\pi^4)c\sig_o(C_V^2+C_A^2)$), and
\beq
\Phi^{\prime}\left(\avemu,\aveamu,p_{\nu_i},p_{\bar{\nu}_i}\right)=
\frac{3}{4}
\left[1-2\avemu\aveamu+p_{\nu_i}p_{\bar{\nu}_i}+
\frac{1}{2}(1-p_{{\nu}_i})(1-p_{\bar{\nu}_i})\right] ,
\eeq
where the flux factor $\avemu$ = $H_\nu/J_\nu $ and the Eddington factor
$p_\nu=\avemut=P_\nu/J_\nu $. Eq. (\ref{jankarate1}) can be rewritten in
terms of the invariant distribution functions $\f_\nu $:
\beq
Q_{\nu_i\anu_i}=
K_i\,\left(\frac{1}{m_ec^2}\right)^2\left(\frac{1}{\hbar c}\right)^6
\int\int\,\Phi^{\prime}\,
\f_{\nu_i}\f_{\bar{\nu}_i}
(\varepsilon_{\nu_i}^4\varepsilon_{\bar{\nu}_i}^3+\varepsilon_{\nu_i}^3\varepsilon_{\bar{\nu}_i}^4)
\,d\varepsilon_{\nu_i}\,d\varepsilon_{\bar{\nu}_i}\, .
\label{jankarate2}
\eeq

Note that when the radiation field is isotropic ($\Phi^{\prime}=1$) and when $\eta_e=0$
the total rate for $e^+e^-$ annihilation given in eq. (\ref{dicusrate2})
equals that for $\nu_i\bar{\nu}_i$ annihilation given in eq.
(\ref{jankarate2}), as expected.  Buras et al. (2002) have addressed the
related and interesting process of $\nu_i\bar{\nu}_i \rightarrow \nu_j\bar{\nu}_j$.
We refer to that paper for a discussion of the relevance and rates of this process.

\section{Nucleon--Nucleon Bremsstrahlung}
\label{bremsst}

A production process for neutrino/anti-neutrino pairs that has recently received 
attention in the supernova context is neutral-current
nucleon--nucleon bremsstrahlung ($n_1 + n_2 \righta n_3 + n_4 + \nu\bar{\nu}$).
It importance in the cooling of old neutron stars, for which the nucleons are quite
degenerate, has been recognized for years (Flowers 1975), but only
in the last few years has it been studied for its potential importance 
in the quasi-degenerate to non-degenerate atmospheres of protoneutron stars
and supernovae (Suzuki 1993; Hannestad \& Raffelt 1998; Burrows et al. 2000; 
Thompson, Burrows, \& Horvath 2000).  Neutron--neutron, proton--proton, and neutron--proton
bremsstrahlung are all important, with the latter the most important for symmetric matter.  As a source of $\nu_e$ and
$\bar{\nu}_e$ neutrinos, nucleon--nucleon bremsstrahlung can not compete
with the charged--current capture processes.
However, for a range of temperatures and densities realized in
supernova cores, it may compete with $e^+e^-$ annihilation as a source
for $\nu_\mu$, $\bar{\nu}_\mu$, $\nu_\tau$, and $\bar{\nu}_\tau$ neutrinos (``$\nu_\mu$''s).
The major obstacles to obtaining accurate estimates of the emissivity of this process are our poor knowledge
of the nucleon--nucleon potential, of the degree of suitability of the Born Approximation, and
of the magnitude of many--body effects (Hannestad \& Rafflet 1998; Raffelt \& Seckel 1998; 
Brinkmann \& Turner 1988).
Since the nucleons in protoneutron star atmospheres are not degenerate,
we present here a calculation of the total and differential
emissivities of this process in that limit and assume a
one-pion exchange (OPE) potential model to calculate the
nuclear matrix element.  For the corresponding calculation for arbitrary nucleon degeneracy,
the reader is referred to \cite{thompson}.  
The formalism we employ has been heavily influenced by those of  
\cite{brinkmann} and \cite{han.raff},
to which the reader is referred for details and further explanations.

Our focus is on obtaining
a useful single--neutrino final--state emission (source) spectrum, as well as a final--state pair energy spectrum
and the total emission rate.  For this, we start with Fermi's Golden Rule for the total rate per neutrino species:
\beqa
Q_{nb}=(2\pi)^4\int \Bigl[\prod_{i=1}^4 \frac{d^3\vec{p}_{i}}{(2\pi)^3}\Bigr]
\frac{d^3\vec{q}_\nu}{(2\pi)^3 2\omega_\nu}
\frac{d^3\vec{q}_{\bar{\nu}}}{(2\pi)^3 2\omega_{\bar{\nu}}}\,
\omega\, \sum_{s}{|{\cal{M}}|}^2 \delta^4({\bf P})\, \Xi_{brems},
\nonumber 
\eeqa
where
\beqa
\Xi_{brems}=\f_1\f_2(1-\f_3)(1-\f_4),
\label{bremfermi}
\eeqa
$\delta^4({\bf P})$ is four--momentum conservation delta function,
$\omega$ is the energy of the final--state neutrino pair,
($\omega_\nu$,$\vec{q}_\nu$) and ($\omega_{\bar{\nu}}$,$\vec{q}_{\bar{\nu}}$)
are the energy and momentum of the neutrino and anti--neutrino, respectively,
and $\vec{p}_{i}$ is the momentum of nucleon $i$.  Final--state
neutrino and anti--neutrino blocking have been dropped.

The necessary ingredients for the integration of eq. (\ref{bremfermi})
are the matrix element for the interaction and a workable procedure for handling
the phase space terms, constrained by the conservation laws.   We follow 
\cite{brinkmann} for both of these elements. In particular, we assume for the
$n + n \righta n + n + \nu\bar{\nu}$ process that the
matrix element is:

\beqa
\sum_{s}{|{\cal{M}}|}^2 = \frac{64}{4} G^2(f/m_\pi)^4 g_A^2 \Bigl[ (\frac{k^2}{k^2+m_{\pi}^2})^2 + \dots \Bigr ]
\frac{\omega_{\nu} \omega_{\bar{\nu}}}{\omega^2}
\nonumber \\
=A\frac{\omega_{\nu} \omega_{\bar{\nu}}}{\omega^2} \, ,
\label{matrixbrem}
\eeqa
where the $4$ in the denominator accounts for the spin average for identical nucleons, $G$ is the
weak coupling constant, $f$ ($\sim1.0$) is the pion--nucleon coupling constant, $g_A$ is the axial--vector
coupling constant, the term in brackets is from the OPE propagator plus exchange and cross terms, $k$ is the nucleon
momentum transfer, and $m_\pi$ is the pion mass.   In eq. (\ref{matrixbrem}), we have dropped $\vec{q}_\nu\cdot\vec{k}$
terms from the weak part of the total matrix element.  To further simplify the calculation, we set the
``propagator'' term equal to a constant $\zeta$, a number of order unity, and absorb into
$\zeta$ all interaction ambiguities.  

Recently, \cite{hanhart} have addressed these
momentum terms in the context of axion emission and $\nu_\mu\bar{\nu}_\mu$
production in supernovae.  In an effort to make contact with the approximation
to the matrix element we present here, they plot $\zeta$ as a function
of average relative thermal nucleon momentum ($\bar{p}$; Phillips, private communication).
The function peaks for $\zeta(\bar{p})$ between $150-200$ MeV at $\zeta\simeq0.47$.
At $\bar{p}=50$ MeV  $\zeta\simeq0.08$ and at $\bar{p}=500$ MeV $\zeta\simeq0.27$.
We are most interested in the region around the $\nu_\mu$ neutrinospheres,
where the emergent spectrum might be most affected by nucleon-nucleon bremsstrahlung.
Mass densities and temperatures in this region might be $10^{12}-10^{13}$ g cm$^{-3}$ and
$5-10$ MeV, respectively.  We estimate $\bar{p}$ in this regime to be $\sim175$ MeV
and take $\zeta=0.5$ for all thermodynamical points. The constant $A$ in eq. (\ref{matrixbrem}) remains.

Inserting a $\int \delta(\omega - \omega_{\nu} - \omega_{\bar{\nu}})d\omega$ by the neutrino phase space terms
times $\omega \omega_{\nu} \omega_{\bar{\nu}}/{\omega^2}$ and integrating over $\omega_{\bar{\nu}}$ yields:

\beq
\int \omega \frac{\omega_{\nu} \omega_{\bar{\nu}}}{\omega^2} \frac{d^3\vec{q}_\nu}{(2\pi)^3
2\omega_\nu}\frac{d^3\vec{q}_{\bar{\nu}}}{(2\pi)^3 2\omega_{\bar{\nu}}}\righta\frac{1}{(2\pi)^4}
\int_{0}^{\infty} \int_{0}^{\omega} \frac{\omega_{\nu}^2 (\omega - \omega_{\nu})^2}{\omega} d\omega_{\nu} d\omega \,  ,
\label{deltaneut}
\eeq
where again $\omega$ equals ($\omega_{\nu} + \omega_{\bar{\nu}}$).  If we integrate
over $\omega_{\nu}$, we can derive the $\omega$ spectrum.  A further integration over $\omega$
will result in the total volumetric energy emission rate.  If we delay such an integration, after
the nucleon phase space sector has been reduced to a function of $\omega$ and if we
multiply eq. (\ref{bremfermi}) and/or eq. (\ref{deltaneut}) by $\omega_{\nu}/\omega$,  an integration
over $\omega$ from $\omega_{\nu}$ to infinity will leave the emission spectrum for the single final--state
neutrino.  This is of central use in multi--energy group transport calculations and
with this differential emissivity and Kirchhoff's Law (\S\ref{stimabs}) we can derive an absorptive opacity.

Whatever our final goal, we need to reduce the nucleon phase space integrals and to do this we use the
coordinates and approach of \cite{brinkmann}.  We define new momenta: $p_+ = (p_1 + p_2)/2$, $p_- = (p_1 - p_2)/2$,
$p_{3c} = p_3 - p_+$, and $p_{4c} = p_4 - p_+$, where nucleons $1$ and $2$ are in the initial state.  Useful direction cosines
are $\gamma_1 = p_+ \cdot p_-/|p_+||p_-|$ and $\gamma_c = p_+ \cdot p_{3c}/|p_+||p_{3c}|$.
Defining $u_i = p_i^2/2mT$ and using energy and momentum conservation, we can show that:
\beqa
d^3p_1d^3p_2 &=& 8d^3p_+d^3p_-
\nonumber \\
\omega &=& 2T(u_- - u_{3c})
\nonumber \\
u_{1,2} &=& u_+ + u_- \pm 2(u_+u_-)^{1/2}\gamma_1
\nonumber \\
u_{3,4} &=& u_{+} + u_{3c} \pm 2(u_+u_{3c})^{1/2}\gamma_c \, .
\label{upm}
\eeqa

In the non--degenerate limit, the $\f_1\f_2(1-\f_3)(1-\f_4)$ term reduces to $e^{2y} e^{-2(u_+ + u_-)}$,
where $y$ is the nucleon degeneracy factor.  Using eq. (\ref{upm}), we see that the quantity $(u_+ + u_-)$ is independent
of both $\gamma_1$ and $\gamma_c$.  This
is a great simplification and makes the angle integrations trivial.
Annihilating $d^3p_4$ with the momentum delta function in eq. (\ref{bremfermi}), noting that $p_i^2dp = \frac{(2mT)^{3/2}}{2}u_i^{1/2}du_i$,
pairing the remaining energy delta function with $u_-$, and integrating $u_+$ from $0$ to $\infty$, we obtain:
\beq
d Q_{nb} = \frac{Am^{4.5}}{2^8\times3\times5 \pi^{8.5}} T^{7.5} e^{2y} e^{-\omega/T}
(\omega/T)^4 \Bigl[ \int_0^{\infty} e^{-x}(x^2 + x\omega/T)^{1/2} dx\Bigr] d\omega \, .
\label{ezz4}
\eeq
The variable $x$ over which we are integrating in eq. (\ref{ezz4}) is equal to $2u_{3c}$.  That integral is analytic and
yields:
\beq
\int_0^{\infty} e^{-x}(x^2 + x\omega/T)^{1/2} dx = \eta e^{\eta}K_1(\eta)\, ,
\label{kintegral}
\eeq
where $K_1$ is the standard modified Bessel function of imaginary argument, related to the Hankel functions, and
$\eta = \omega/2T$.  Hence, the $\omega$ spectrum is given by:
\beq
\frac{d Q_{nb}}{d\omega} \propto e^{-\omega/2T} \omega^5 K_1(\omega/2T) \, .
\label{omegaspect}
\eeq

It can easily be shown that $\langle \omega \rangle = 4.364 T$.
Integrating eq. (\ref{ezz4}) over $\omega$ and using the thermodynamic identity in the non--degenerate limit:
\beq
e^y = \Bigl(\frac{2\pi}{mT}\Bigr)^{3/2} n_n/2 \, ,
\eeq
where $n_n$ is the density of neutrons (in this case), we derive for the
total neutron--neutron bremsstrahlung emissivity of a single neutrino pair:
\beq
Q_{nb} = 1.04\times10^{30}
\zeta(X_n \rho_{14})^2 (\frac{T}{{\rm MeV}})^{5.5} \, {\rm ergs\, cm^{-3}\, s^{-1}} \, ,
\label{bremssr}
\eeq
where $\rho_{14}$ is the mass density in units of $10^{14}$ gm cm$^{-3}$ and
$X_n$ is the neutron mass fraction.  Interestingly,  this is
within 30\% of the result in \cite{suzuki_93}, even though he has substituted, without much justification, $(1+\omega/2T)$ for
the integral in eq. (\ref{ezz4}). ($[1+(\pi\eta/2)^{1/2}]$ is a better
approximation.)  The proton--proton and neutron--proton processes can be handled similarly and the total
bremsstrahlung rate is then obtained by substituting $X_n^2 + X_p^2 + \frac{28}{3} X_n X_p$ for
$X_n^2$ in eq. (\ref{bremssr}) (Brinkmann and Turner 1988).
At $X_n = 0.7$, $X_p = 0.3$, $\rho = 10^{12}$ gm cm$^{-3}$, and T = 10 MeV, and taking the
ratio of augmented eq. (\ref{bremssr}) to eq. (\ref{enumurate}),
we obtain the promising ratio of $\sim 5\zeta$.
Setting the correction factor $\zeta$ equal to $\sim0.5$ (Hanhart, Phillips, and Reddy 2001), we find
that near and just deeper than the $\nu_\mu$ neutrinosphere, bremsstrahlung is larger than classical
pair production.

If in eq. (\ref{deltaneut}) we do not integrate over $\omega_\nu$, but at the
end of the calculation we integrate over $\omega$ from $\omega_\nu$ to $\infty$,
after some manipulation we obtain the single neutrino emissivity spectrum:
\beqa
\frac{d Q_{nb}^{\prime}}{d\omega_{\nu}} =
2C \Bigl(\frac{Q_{nb}}{T^4}\Bigr)
\omega_{\nu}^3 \int^{\infty}_{\eta_\nu}  \frac{e^{-\eta}}{\eta} K_1(\eta) (\eta - {\eta_\nu})^2 d\eta
\eeqa
\beqa
= 2C \Bigl(\frac{Q_{nb}}{T^4}\Bigr)
\omega_{\nu}^3 \int^{\infty}_{1} \frac{e^{-2\eta_{\nu}\xi}}{\xi^3} (\xi^2-\xi)^{1/2} d\xi \, ,
\label{spectrum}
\eeqa
where $\eta_{\nu} = \omega_\nu/2T$, $C$ is the normalization constant equal
to $\frac{3\times5\times7\times11}{2^{11}}$ ($\cong 0.564$), and for the second expression we have used the
integral representation of $K_1(\eta)$ and reversed the order of integration.  In eq. (\ref{spectrum}),
$Q_{nb}$ is the emissivity for the pair.

Eq. (\ref{spectrum}) is the approximate neutrino emission spectrum  due to nucleon--nucleon bremsstrahlung.
A useful fit to eq. (\ref{spectrum}), good to better than 3\% over the full range of important values of $\eta_{\nu}$, is:
\beq
\frac{d Q_{nb}^{\prime}}{d\omega_{\nu}} \cong
\frac{0.234 Q_{nb}}{T} \Bigl(\frac{\omega_\nu}{T}\Bigr)^{2.4} e^{-1.1 \omega_{\nu}/T}\, .
\eeq
Thompson, Burrows, and Horvath (2000) should be consulted for a detailed discussion 
of nucleon-nucleon bremsstrahlung for arbitrary nucleon 
degeneracy.

\section{Conclusion}
 
\begin{figure}
\vspace*{6.0in}
\hbox to\hsize{\hfill\includegraphics{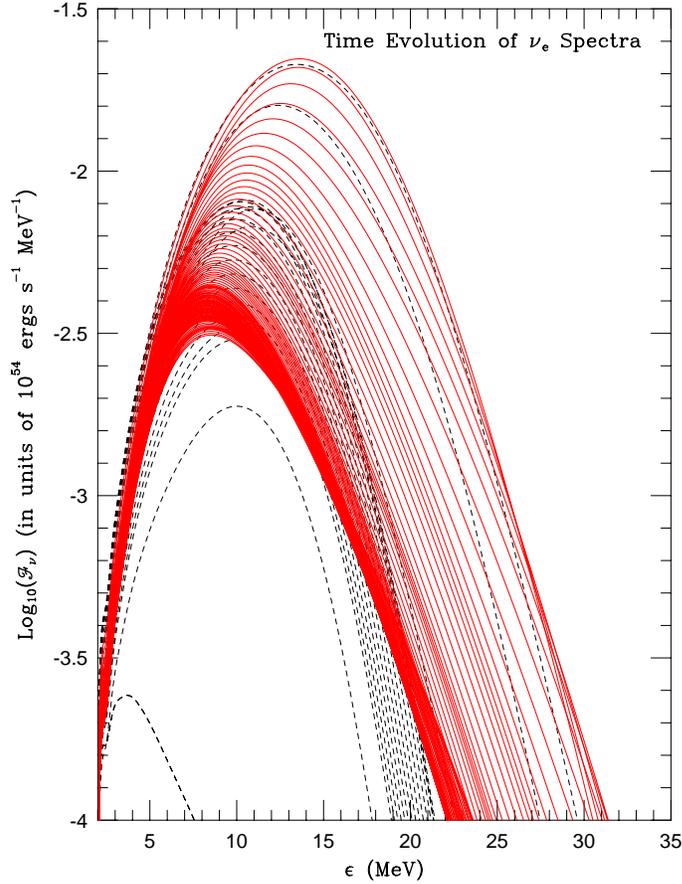}\kern+6in\hfill}
\caption{A collection of emergent $\nu_e$ spectra at various times 
during the core-collapse, bounce, and shock-stagnation phases 
of the core of an 11 M$_{\odot}$ progenitor.  The luminosity spectrum
(logarithm base ten) is in units of 10$^{54}$ ergs s$^{-1}$ MeV$^{-1}$
and the neutrino energy (abscissa) is in units of MeV.  The dashed curves 
cover the collapse phase (of duration $\sim$200 milliseconds) until just before the peak luminosity
around shock breakout is achieved and the solid curves are for the subsequent
cooling and deleptonization phases after the peak.  The last curve is at
$\sim$110 milliseconds after bounce.  The lowest dashed curve
is at a time early during collapse. Note the relative softness of the spectrum
then.  As the figure shows, the transition from the dashed to the solid curve 
happens close to the time when the $\nu_e$ spectrum is hardest.
}
\label{specte}
\end{figure}

\begin{figure}
\vspace*{6.0in}
\hbox to\hsize{\hfill\includegraphics{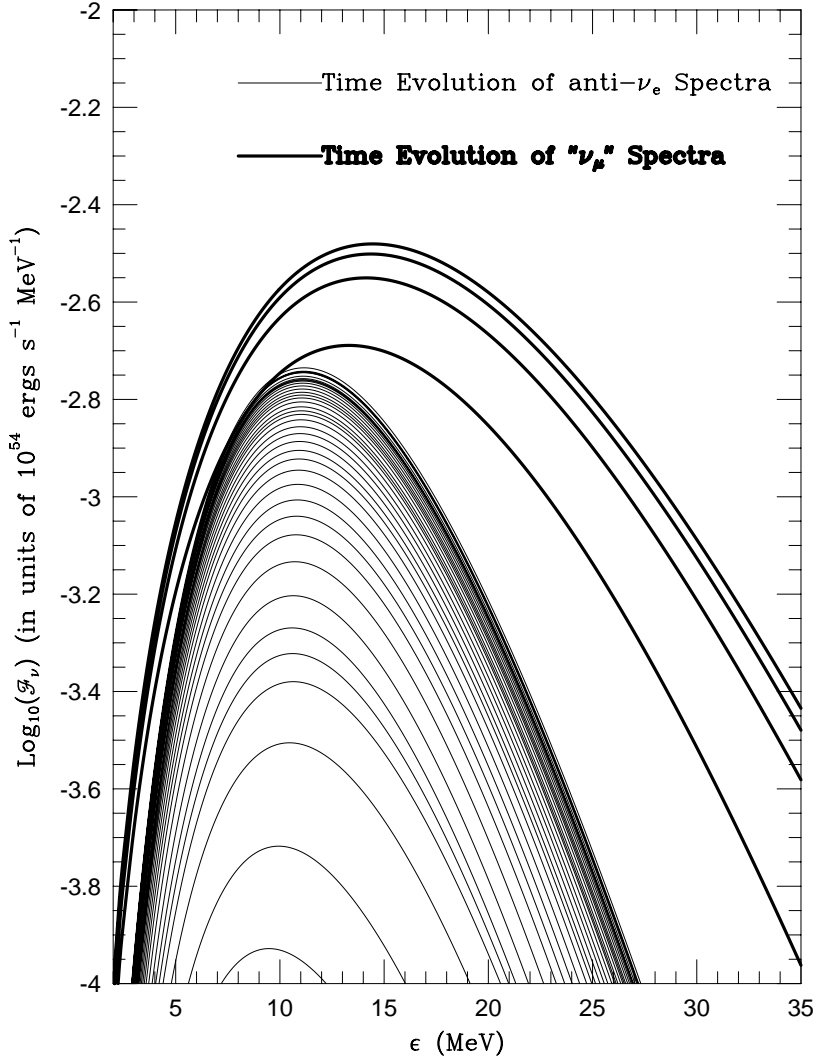}\kern+6in\hfill}
\caption{This figure shows the $\bar{\nu}_e$ (thin lines)
and ``$\nu_{\mu}$'' (thick lines) emergent luminosity spectra for the 11 M$_{\odot}$
progenitor evolution depicted in Fig. \ref{specte}.  The luminosity spectra
(logarithm base ten) are in units of 10$^{54}$ ergs s$^{-1}$ MeV$^{-1}$
and the neutrino energy (abscissa) is in units of MeV. There is no appreciable flux prior to shock breakout
for these species.  To avoid clutter, we here depict only a few $\nu_{\mu}$ spectra
to $\sim$50 milliseconds after bounce.  (These curves represent the sum of 
the $\nu_{\mu}$, $\bar{\nu}_{\mu}$, $\nu_{\tau}$, and $\bar{\nu}_{\tau}$ 
luminosity spectra.)  However, the $\bar{\nu}_e$ spectra are shown until about
110 milliseconds after bounce.  During the phases shown, both sets of luminosities are always
increasing.  Note that the $\nu_{\mu}$ spectra are significantly harder than
either the $\bar{\nu}_e$ or the $\nu_e$ spectra.  This is a consequence of 
of the fact that the $\nu_{\mu}$s do not have appreciable charged-current
cross sections (eqs. \ref{ncapture} and \ref{pcapture}), enabling one to probe more
deeply into the hot core with these species.   
}
\label{spectam}
\end{figure}

The processes that have been described above are essential elements  
of the neutrino-driven supernova explosion mechanism.  Coupling these with radiation-hydrodynamics
codes, an equation of state, beta-decay and electron capture microphysics, and
nuclear rates, one explores the viability of various scenarios for the explosion
of the cores of massive stars (Liebend\"orfer et al. 2001ab; 
Rampp \& Janka 2000).  Recently, Thompson, Burrows, and Pinto (2002) have 
incorporated this neutrino microphysics into simulations of 1D (spherical) core collapse and
have investigated the effects on the dynamics, luminosities, and emergent 
spectra of weak magnetism/recoil, nucleon-nucleon bremsstrahlung, inelastic
neutrino-electron scattering, and a host of the cross section corrections described
above.  Figures \ref{specte} and \ref{spectam} depict some of the resultant
luminosity spectra and their temporal evolution for a representative simulation.  
The character of the spectra
reflect the opacities and sources described in this paper.  In particular, the
energy hardness hierarchy from $\nu_e$ (softer) to $\nu_{\mu}$ (harder) neutrinos
is clearly demonstrated on these plots, as is the distinction between the $\nu_e$ pre-breakout
and post-breakout spectra (Fig.~\ref{specte}). (See the figure captions for further details.)

To date, none of the detailed 1D simulations that have been performed 
explodes and it may be that multi-dimensional effects play a pivotal role in the explosion mechanism 
(Herant et al.~1994; Burrows, Hayes, \& Fryxell 1995; Fryer et al. 1999; Janka \& M\"uller 1996).
Be that as it may, an understanding of neutrino-matter interactions remains
central to unraveling one of the key mysteries of the nuclear universe in which
we live.  

\acknowledgments

We would like to thank Ray Sawyer, Sanjay Reddy, and Jorge Horvath for fruitful
discussions and/or collaboration on some of the more thorny 
aspects of neutrino-matter interactions.  
Support for this work was provided by
the Scientific Discovery through Advanced Computing (SciDAC) program
of the DOE, grant number DE-FC02-01ER41184, and
by NASA through Hubble Fellowship
grant \#HST-HF-01157.01-A awarded by the Space Telescope Science
Institute, which is operated by the Association of Universities for
Research in Astronomy, Inc., for NASA, under contract NAS 5-26555.

\end{document}